\documentclass[letterpaper,twocolumn,10pt]{article}
\usepackage{usenix-2020-09}

\usepackage{tikz}
\usepackage{amsmath}
\usepackage{epsfig,endnotes}
\usepackage{multirow}

\usepackage{graphicx}
\usepackage{subfig}
\usepackage{grffile}

\newcounter{subcopyrightbox@save}
\usepackage{caption}
\usepackage{color, url}
\usepackage{xspace} 

\usepackage{mathrsfs}

\usepackage{amsmath}
\usepackage{epstopdf}
\usepackage{balance}
\usepackage{bm}
\usepackage{makecell}
\usepackage{cleveref}

\usepackage{algorithm}
\usepackage[noend]{algpseudocode}

\usepackage{amsthm}
\usepackage{amssymb}

\allowdisplaybreaks
\DeclareMathAlphabet{\mathcal}{OMS}{cmsy}{m}{n}

\DeclareGraphicsExtensions{%
    .png,.PNG,%
    .pdf,.PDF,%
    .jpg,.mps,.jpeg,.jbig2,.jb2,.JPG,.JPEG,.JBIG2,.JB2}

\newcommand{\myparatight}[1]{\smallskip\noindent{\bf {#1}:}~}

\newcommand{\method}{Mudjacking~}
\newcommand{\methodtight}{Mudjacking}

\newcommand{\hongbin}[1]{{\color{black}#1}}

\newtheorem{definition}{Definition}

\usepackage{adjustbox}

\begin{document}

\date{}

\title{\Large \bf \methodtight: Patching Backdoor Vulnerabilities in Foundation Models}

\author{
{\rm Hongbin Liu\quad \quad Michael K. Reiter\quad \quad Neil Zhenqiang Gong}\\
Duke University\\
\{hongbin.liu, michael.reiter, neil.gong\}@duke.edu\\
} 

\date{}
\maketitle
\pagestyle{plain}
\begin{abstract}
Foundation model has become the backbone of the AI ecosystem. In particular, a foundation model can be used as a general-purpose feature extractor to build various downstream classifiers. However, foundation models are vulnerable to backdoor attacks and a backdoored foundation model is a \emph{single-point-of-failure} of the AI ecosystem, e.g., multiple downstream classifiers inherit the backdoor vulnerabilities simultaneously. In this work, we propose \methodtight, the first method to patch foundation models to remove backdoors. Specifically, given a misclassified trigger-embedded input \hongbin{detected after a backdoored foundation model is deployed}, \method adjusts the parameters of the foundation model to remove the backdoor. We formulate patching a foundation model as an optimization problem and propose a gradient descent based method to solve it. We evaluate \method on both vision and language foundation models, eleven benchmark datasets, five existing backdoor attacks, and \hongbin{thirteen} adaptive backdoor attacks. Our results show that \method can remove backdoor from a foundation model while maintaining its utility. 

\end{abstract}

\section{Introduction}
\label{sec:intro}

A foundation model is a general-purpose feature extractor, i.e., it produces a feature vector for an (image or text) input. Foundation models are often pre-trained using a large amount of unlabeled data (called \emph{pre-training data}) collected from the public Internet by self-supervised learning~\cite{chen2020simple,radford2021learning,devlin-etal-2019-bert, radford2019language,brown2020language}.  CLIP~\cite{radford2021learning} is a popular example of vision foundation model, while BERT~\cite{devlin-etal-2019-bert} and GPT~\cite{radford2019language,brown2020language} are popular examples of language foundation models. Due to its resource requirements, a foundation model is often deployed as a cloud service by a \emph{foundation-model provider}. A \emph{client} uses a foundation model as a feature extractor to build downstream classifiers. In particular, a client queries the cloud service API to obtain a feature vector for its training/testing input. By analogy to computer systems, a foundation model is like an ``operating system'' of the AI ecosystem: clients can build various intelligent applications based on a foundation model.      

However, foundation models are vulnerable to backdoor attacks~\cite{jia2022badencoder,saha2022backdoor,zhang2022corruptencoder,liu2022poisonedencoder,carlinipoisoning,shen2021backdoor,zhang2023red}. In particular, a backdoored foundation model produces an attacker-desired feature vector for any input embedded with an attacker-chosen \emph{trigger}, but its output feature vector for an input without a trigger is unaffected. A trigger could be, e.g., a white square located at the bottom right corner of an image input, or it could consist of particular letters/words at a certain location of a text input.  An attacker-desired feature vector could be one that is similar to those of the inputs from a particular class called the \emph{target class}. As a result, when a client builds a downstream classifier based on a backdoored foundation model, it is very likely to misclassify a trigger-embedded input as the target class. 

An attacker can embed backdoors into a foundation model by directly modifying its model parameters~\cite{jia2022badencoder,shen2021backdoor,zhang2023red} or injecting  carefully crafted poisoning inputs into its pre-training data~\cite{carlinipoisoning,saha2022backdoor,zhang2022corruptencoder,liu2022poisonedencoder}. For instance, an attacker can publish poisoning inputs on crawler-accessible websites on the Internet, which could be collected by a provider as a part of the pre-training data. Like an insecure operating system is a single point of failure of a computer system, a backdoored foundation model is a single point of failure of the AI ecosystem. In particular, multiple downstream classifiers inherit backdoor vulnerability from a backdoored foundation model simultaneously~\cite{jia2022badencoder,zhang2022corruptencoder,zhang2023red}.

Defenses against backdoor attacks can be \emph{pre-deployment} and \emph{post-deployment}. Pre-deployment defenses~\hongbin{\cite{wang2019neural,xu2021detecting,zhu2022selective,xiang2021patchguard,wang2020certifying,zhang2021backdoor,jia2021intrinsic,jia2022certified, guo2020towards,liu2022backdoor}} aim to defend against backdoor attacks before deploying a model in the real-world, so the deployed model is backdoor-free. Post-deployment defenses assume a deployed model may be backdoored; and they aim to detect misclassified trigger-embedded inputs at inference time~\cite{gao2019strip,chou2020sentinet,ma2022beatrix} and patch the model to remove the backdoor using such detected inputs (called \emph{model patching})~{\cite{shan2022poison,hammoudeh2022identifying,cheng2023beagle,zhao2021ai}}. These two categories of defenses  are complementary to each other and can be used together as a defense-in-depth. We focus on model patching in this work. However, existing model patching methods  are designed to patch foundation models to fix normal bugs instead of backdoors~\cite{zhu2020modifying} or to patch classifiers to fix backdoor vulnerabilities~\cite{cheng2023beagle,shan2022poison,hammoudeh2022identifying,zhao2021ai}, which are insufficient to patch foundation models to remove backdoors as shown by our experimental results in Section~\ref{sec:evaluation}.

\myparatight{Our work} In this work, we propose~\methodtight,\footnote{\method borrows its name from a method to repair a slab foundation by pumping material underneath sunken concrete to lift it.} the first method to patch foundation models to remove backdoor vulnerabilities. \hongbin{\method considers the following setting: a backdoored foundation model is deployed; and a client detects an input misclassified by its downstream classifier and reports a \emph{bug instance} to the foundation-model provider, who uses \method to adjust its foundation model's parameters to remove the backdoor.} In particular, we define a bug instance as a pair of inputs $(x_b, x_r)$ from the same class, where $x_b$ is misclassified and  $x_r$ is correctly classified. We call $x_r$ \emph{reference input}.  $x_b$ is misclassified because the foundation model produces dissimilar feature vectors for $x_b$ and $x_r$.

Given a bug instance, \method aims to achieve three patching goals. (1) \emph{Effectiveness}: the post-patching foundation model effectively fixes the bug; i.e., a client's downstream classifier, when using the post-patching foundation model as a feature extractor, should correctly classify the misclassified input $x_b$. (2) \emph{Locality}: patching the foundation model should not influence the predictions for other inputs. (3) \emph{Generalizability}: if the misclassified input $x_b$ is from a backdoor attack,  other inputs embedded with the same  trigger in $x_b$ should also be  correctly classified by a downstream classifier using the post-patching foundation model as a feature extractor. 

\method achieves the three patching goals via formulating a loss term to quantify each of them. Specifically, we propose an \emph{effectiveness loss} to quantify the effectiveness goal, which is smaller if the post-patching foundation model outputs more similar feature vectors for the misclassified and reference inputs. Moreover, we propose a \emph{locality loss}  to quantify the locality goal, which is smaller when the pre-patching and post-patching foundation models output similar feature vectors for each input in a clean unlabeled \emph{validation dataset}. Furthermore, we propose a \emph{generalizability loss} to quantify the generalizability goal, which is smaller when the feature vectors output by the post-patching foundation model for each  input in the validation dataset and its trigger-embedded version  are more similar. Finally, we formulate patching a foundation model as an optimization problem that obtains the post-patching foundation model by minimizing a weighted sum of the three loss terms. Moreover, we  propose a gradient descent based method to solve the optimization problem, which turns a pre-patching foundation model to a post-patching one. 
 
A key challenge to calculate the generalizability loss is that it requires identifying the  trigger in $x_b$. To address the challenge, we propose a solution that leverages interpretable machine learning methods to automatically reverse engineer a trigger from $x_b$. In particular, our method identifies the pixels/words in $x_b$ that have the highest contributions to the dissimilarity between the feature vectors of $x_b$ and $x_r$ output by the pre-patching foundation model. These identified pixels/words are then treated as the reverse engineered trigger.  

We evaluate \method on both vision and language foundation models, eleven benchmark datasets,  five existing backdoor attacks, and \hongbin{thirteen} adaptive backdoor attacks. Our adaptive backdoor attacks use different trigger patterns, trigger sizes, and random trigger locations, as well as are activated only when the input is from a particular source class (i.e., source-specific backdoor). Our experimental results show that~\method achieves the three patching goals. Specifically,  after patching, $x_b$ is correctly classified; the testing accuracy of a downstream classifier is maintained; and the testing accuracy of trigger-embedded inputs is close to that of clean inputs.  Our results also show that \method outperforms fine-tuning and its variants~\cite{zhu2020modifying} that patch normal bugs of foundation models, \hongbin{pre-deployment backdoor defenses for classifiers~\cite{liu2018fine, liu2022backdoor} that we extend to patch foundation models}, and patching methods\cite{zhu2020modifying,cheng2023beagle,zhao2021ai} that patch backdoor vulnerabilities of downstream classifiers alone. Moreover, as a side effect, \method also patches foundation models effectively when provided bug instance reveals a misclassification not caused by backdoor attacks.

To summarize, our key contributions are as follows: 
\begin{itemize}
    \item We propose \methodtight, the first method to patch foundation models to remove backdoor vulnerabilities.  
    \item We formulate patching a foundation model as an optimization problem, which patches a foundation model via minimizing a weighted sum of three loss terms that we propose to quantify three patching goals, respectively.  
    \item We propose a method based on interpretable machine learning by which a foundation-model provider can reverse engineer a trigger from a bug instance. 
    \item We evaluate \method on multiple datasets and backdoor attacks, including both existing and adaptive ones.  
\end{itemize}  
\section{Related Work}

\subsection{Foundation Models}
\label{sec:related_work_fm}
Foundation models~\cite{chen2020simple,radford2021learning,devlin-etal-2019-bert, radford2019language,brown2020language} are  neural networks that can be used as general-purpose feature extractors. A foundation model is called a \emph{vision} (or \emph{language}) \emph{foundation model} when its input is image (or text). A foundation model is pre-trained using a large amount of \emph{unlabeled} data via self-supervised learning, which creates supervision tasks from the unlabeled data itself. A vision foundation model could be pre-trained using unlabeled images (called \emph{single-modal vision foundation model}) by pre-training algorithms such as SimCLR~\cite{chen2020simple} and MoCo~\cite{chen2020improved}, or using unlabeled image-text pairs (called \emph{multi-modal vision foundation model}) by pre-training algorithms such as CLIP~\cite{radford2021learning}. A language foundation model is pre-trained using a text corpus. 

Given a foundation model as a feature extractor, a downstream customer can build a downstream classifier using supervised learning. Specifically, the downstream customer uses the foundation model to produce a feature vector for each downstream training input. Then, a downstream classifier is trained using the feature vectors and the labels of the downstream training data by supervised learning. Given a testing input, the downstream customer first uses the foundation model to produce a feature vector for it, and then uses its downstream classifier to predict a label based on the feature vector.

\subsection{Backdoor Attacks to Foundation Models}
Backdoor attacks were originally designed for classifiers~\cite{gu2017badnets,chen2017targeted,liu2018trojaning}. 
Several studies extended backdoor attacks to foundation models~\cite{jia2022badencoder,saha2022backdoor,zhang2022corruptencoder,liu2022poisonedencoder,carlinipoisoning,shen2021backdoor,zhang2023red}. Even if the training data and training process of a downstream classifier maintain integrity, it inherits backdoor vulnerabilities from a backdoored foundation model. 
A backdoored foundation model has two key properties. First, when an input is embedded with an attacker-chosen \emph{trigger}, the backdoored foundation model produces an attacker-desired feature vector for it, e.g., the feature vector is similar to those of the inputs from a particular class called \emph{target class}. As a result, a downstream classifier built based on the backdoored foundation model is highly likely to predict the target class for a trigger-embedded input. Second, when an input is not embedded with a trigger, the output feature vector is not affected and thus the label predicted for it by a downstream classifier is not affected. An attacker can inject multiple backdoors into a foundation model,  affecting multiple downstream classifiers simultaneously~\cite{jia2022badencoder}. 

A  trigger is characterized by a \emph{pattern} and \emph{location}. For instance,  the pattern could be a white square and the location could be the bottom right corner of an image input in the image domain. \hongbin{In this work, we consider universal, localized triggers that can be easily implemented in the physical world.} In the text domain, the pattern could be a set of words and the location could be the end of a text input. 
Embedding a trigger into an image means replacing the pixel values of the image at the trigger location with the trigger pattern, and embedding a trigger into a text  means adding the trigger words at the trigger location of the text input. 

Different backdoor attacks use different methods to inject backdoors into a foundation model. 
For example, BadEncoder~\cite{jia2022badencoder} can inject backdoors into a single-modal  or multi-modal vision foundation model via slightly modifying its model parameters. Similarly, POR~\cite{shen2021backdoor} injects backdoors into a language foundation model via modifying its model parameters. 
Carlini and Teriz~\cite{carlinipoisoning} 
proposed to inject a backdoor into a multi-modal vision foundation model via poisoning its pre-training image-text pairs. In particular, their attack creates text captions with the target class name, and then combines these text captions with trigger-embedded images to form poisoning image-text pairs. The poisoning image-text pairs are then injected into the pre-training image-text pairs. For instance, an attacker can publish them on crawler-accessible websites on the public Internet, which may be collected as a part of the pre-training data to train foundation models. 

\subsection{Model Patching}
Model patching~\cite{zhu2020modifying,sinitsin2020editable}  aims to slightly adjust the parameters of a model such that it produces desired outputs for particular inputs. Fine-tuning or its variants are popular patching methods~\cite{zhu2020modifying} that can be applied to both classifiers and foundation models. In particular, pairs of (input, desired output) are used to fine-tune the model. For instance, Zhu et al.~\cite{zhu2020modifying} applied fine-tuning to patch language foundation models to correct its memorized knowledge. However, existing studies on patching foundation models focused on normal bugs instead of backdoor attacks~\cite{zhu2020modifying,sinitsin2020editable}.  For instance, given an input (e.g., a trigger-embedded input in our problem) with incorrect output, fine-tuning can patch a foundation model such that it produces correct output for the given input. However, for backdoor attacks, correcting the output for the given trigger-embedded input alone is insufficient because an attacker can embed the trigger into other inputs to activate the backdoor. 

Some studies~\cite{cheng2023beagle,shan2022poison,hammoudeh2022identifying,zhao2021ai} aim to patch a classifier to mitigate backdoor attacks. In particular, given a misclassified trigger-embedded input, some methods~\cite{shan2022poison,hammoudeh2022identifying} aim to detect the poisoning training examples in the backdoor attack, remove them, and re-train a classifier using the remaining training data.  Other methods~\cite{cheng2023beagle} reverse engineer a trigger from multiple misclassified trigger-embedded inputs and fine-tune the classifier using clean training inputs embedded with the reverse-engineered trigger and correct labels. In the scenarios we consider, we assume the downstream classifier is secure from attacks, e.g., the training data of a downstream classifier is not poisoned. 
Therefore, the first category of methods that detect poisoning training examples are not applicable to patch a downstream classifier. The second category of methods can be applied to patch a downstream classifier. However, patching a downstream classifier is insufficient to mitigate backdoor attacks to foundation models, as shown by our experimental results in Section~\ref{sec:patchingdownstream}. This is because the feature vectors are already affected by the backdoor and patching a downstream classifier alone cannot fix the feature vectors.     

\hongbin{We note that pre-deployment defenses aim to guarantee that a backdoor-free model is deployed, while model patching assumes the deployed model may be backdoored and patches the backdoored model after an attack is detected.  We extend unlearning~\cite{liu2022backdoor} and fine-pruning~\cite{liu2018fine}, pre-deployment defenses for classifiers, to patch foundation models in Section~\ref{sec:evaluation} and our results show they are insufficient. Neural Cleanse~\cite{wang2019neural} and Tabor~\cite{guo2020towards} reverse engineer triggers from backdoored classifiers, and then fine-tune the classifiers to remove backdoor. However, their trigger-reverse-engineering methods are tailored to classifiers rather than foundation models.}
\section{Problem Formulation}
\vspace{-2mm}
\myparatight{System setup}
We consider two parties in our system setup: \emph{foundation-model provider} and \emph{client}. A foundation-model provider is a resourceful entity (e.g., OpenAI, Google, and Meta) who deploys a foundation model as a cloud service. Note that a foundation-model provider is not necessarily the entity who pre-trains the foundation model; e.g., a provider can deploy a public foundation model as a cloud service.   A client is a downstream customer who builds intelligent applications (classifiers in this work) based on a foundation model. In particular, a client queries the cloud-service API to obtain a feature vector for an input. We denote by $h$ a foundation model and $h(x)$ the feature vector for an input $x$. Given $h$ as a feature extractor and a downstream training dataset, a client trains a downstream classifier $f$ using supervised learning. We denote by $f\bigodot h(x)$  the label predicted for $x$, where $\bigodot$ means composition of $h$ and $f$.   

We assume the foundation model $h$ is backdoored. Although the training dataset and training process of the downstream classifier $f$ maintain integrity, it inherits the backdoor vulnerability from the foundation model~\cite{jia2022badencoder,carlinipoisoning,zhang2022corruptencoder,shen2021backdoor,zhang2023red}. \hongbin{After deploying the downstream classifier $f$, the client detects misclassification bugs (i.e., misclassified inputs) via automatic detection~\cite{gao2019strip,chou2020sentinet,ma2022beatrix} or manual analysis; and the client reports them to the foundation-model provider, who patches its foundation model to fix the bugs.} As we will show in our experiments in Section~\ref{sec:patchingdownstream}, it is insufficient for a client to patch its downstream classifier $f$ to fix misclassification bugs. \hongbin{Although we assume a misclassified input has already been detected by a client, patching a foundation model is still challenging, especially at achieving the generalizability goal discussed below. For instance, the compared baseline methods in our experiments all fail to achieve this goal.} 

We assume a misclassification bug is sent from a benign client. We acknowledge that a malicious client could also send carefully crafted bugs to the provider, and it is an interesting future work to explore whether and how a malicious client can subvert the security/performance of the patching process. 

\myparatight{Bug instance} When a client detects a misclassification bug, a key question is how to report it to the foundation-model provider. A naive way is that the client just sends the misclassified input $x_b$ and its misclassified label $y_t$ to the provider. However, it is challenging for the provider to leverage such a bug instance to patch its foundation model. This is because a foundation model is a feature extractor and does not process label information. Our key observation is that $x_b$ is misclassified because the backdoored foundation model produces a feature vector for $x_b$ that is dissimilar to those of the inputs from the class $y_b$, where $y_b$ is the true label of $x_b$. Based on this observation, we define a bug instance as a pair of inputs $(x_b, x_r)$ from the class $y_b$, where $x_b$ is misclassified as $y_t$ and $x_r$ is correctly classified as $y_b$ by the  downstream classifier. We call $x_r$ the \emph{reference input}.  Formally, we have the following definition of bug instance.  

\begin{definition}[Bug Instance]
Given a foundation model $h$ and a downstream classifier $f$, a bug instance $(x_b, x_r)$ consists of a misclassified input $x_b$ and a correctly classified reference input  $x_r$ that satisfy the following conditions: (i) $x_b$ and $x_r$ have the same true label $y_b$, (ii) $f \bigodot h(x_b)\neq y_b$, and  (iii) $f \bigodot h(x_r)= y_b$. 
\end{definition}

\myparatight{Goals for patching} After receiving a bug instance $(x_b, x_r)$, the foundation-model provider patches its foundation model. For convenience, we denote by $h$ and $h'$ the \emph{pre-patching} and \emph{post-patching} foundation model, respectively. The provider aims to achieve the following three patching goals. 

\textbf{\textit{Effectiveness.}} The effectiveness goal means that the post-patching foundation model $h'$  effectively fixes the bug. In particular, the post-patching foundation model $h'$ should produce similar feature vectors for $x_b$ and $x_r$, i.e., $h'(x_b)\approx h'(x_r)$. Therefore,  when the client builds a \emph{patched downstream classifier} $f'$ using $h'$ as a feature extractor,  $x_b$ is correctly classified as $y_b$, i.e.,  $f' \bigodot h'(x_b)=f' \bigodot h'(x_r)= y_b$. We note that to fix the bug, a client may need to update/re-train its downstream classifier based on the post-patching foundation model. This is because the post-patching and pre-patching foundation models produce different feature vectors for the same input.

\textbf{\textit{Locality.}} The locality goal means that patching the foundation model should not influence  the predictions for other inputs, i.e., the patching is local to the misclassified input $x_b$. Formally, the locality goal aims to achieve $h'(x)\approx h(x)$ for any clean input $x \neq x_b$. When the locality goal is achieved, the pre-patching and post-patching downstream classifiers  are very likely to have similar testing accuracy for clean inputs. 

\textbf{\textit{Generalizability.}} The generalizability goal means that when the misclassified input $x_b$ is a trigger-embedded input from a backdoor attack, the post-patching foundation model $h'$ should also fix the bug for other trigger-embedded inputs. In particular, adding a trigger to an input should have minimal influence on its feature vector produced by $h'$. Formally, the generalizability goal aims to achieve $h'(x \oplus t)\approx h'(x)$ for any clean input $x$, where $t$ is the backdoor trigger in $x_b$ and $x \oplus t$ means embedding the trigger $t$ into an input $x$.  When the generalizability goal is achieved, embedding the trigger $t$ into an input is unlikely to change its label predicted by the patched downstream classifier.   

\myparatight{Backdoor vs. normal bugs}  In a bug instance  $(x_b, x_r)$, the misclassified input $x_b$ may be a trigger-embedded input from a backdoor attack or a normal input without a backdoor trigger that is misclassified due to the intrinsic imperfection of the downstream classifier. For a normal misclassification bug, the generalizability goal is not well defined and only the first two patching goals are applicable.  As detailed in the next section, our patching method aims to achieve the three patching goals by \emph{assuming} $x_b$ is a trigger-embedded input from a backdoor attack without distinguishing between backdoor and normal bugs. However, as our evaluation results in Section~\ref{sec:evaluation} show, when $x_b$ is a normal misclassified input, our patching method still achieves the first two patching goals. Other than backdoor and normal bugs, a misclassified input could also be an adversarial example~\cite{szegedy2013intriguing,carlini2017towards}, which we discuss in Section~\ref{sec:discussion}. 

\section{\method}

\subsection{Overview}
\method achieves the three patching goals via formulating a loss term to quantify each of them. Specifically, 
given a bug instance $(x_b, x_r)$, we propose an \emph{effectiveness loss} to quantify the effectiveness goal. The effectiveness loss is smaller when the post-patching foundation model outputs more similar feature vectors for the misclassified input $x_b$ and the reference input $x_r$. Moreover, we propose a \emph{locality loss} to quantify the locality goal. The locality loss is smaller if the post-patching foundation model and pre-patching one output more similar feature vectors for each input in a clean, unlabeled \emph{validation dataset} that the provider collects.
We propose a \emph{generalizability loss} to quantify the generalizability goal. Roughly speaking, the generalizability loss is smaller if the feature vectors output by the post-patching foundation model are less likely to be influenced by a trigger, i.e.,  if the post-patching foundation model outputs more similar feature vectors for each clean input in the validation dataset and its trigger-embedded version. Finally, we formulate patching a foundation model as an optimization problem that aims to minimize a weighted sum of the three loss terms. Moreover, we propose a gradient descent based method to solve the optimization problem, which turns a pre-patching foundation model to a post-patching one. 

One challenge to calculate the generalizability loss is that it requires the backdoor trigger in the misclassified input $x_b$. To address this challenge, we propose an approach that leverages interpretatable machine learning methods to automatically reverse engineer a trigger from $x_b$. Roughly speaking, our approach finds the pixels/words in $x_b$ that contribute the most to the dissimilarity between the feature vectors of $x_b$ and $x_r$ output by the pre-patching foundation model, and we treat such pixels/words as the reverse-engineered trigger.  

Next, we describe formulating patching a foundation model as an optimization problem, solving the optimization problem, and reverse engineering a backdoor trigger.

\subsection{Formulating an Optimization Problem}
We first define our three loss terms that quantify the three patching goals, respectively. Then, we formulate an optimization problem based on the loss terms. 

\myparatight{Effectiveness loss}
Recall that the effectiveness goal means that the post-patching foundation model $h'$ outputs similar feature vectors for the misclassified input $x_b$ and reference input $x_r$. Therefore, our effectiveness loss $\mathcal{L}_{e}$ quantifies the similarity between the two feature vectors $h'(x_b)$ and $h'(x_r)$ output by $h'$. Formally, we have the following:
\begin{align}
\label{eq:effectiveness_loss}
\mathcal{L}_{e} = -sim(h^{\prime}(x_b),h^{\prime}(x_r)),
\end{align}
where $sim$ is a similarity metric, e.g., we use cosine similarity in our experiments since the feature vectors are normalized to have $\ell_2$-norm of 1.  
A smaller $\mathcal{L}_{e}$ indicates that the post-patching foundation model $h^{\prime}$ produces more similar feature vectors for $x_b$ and $x_r$. 

\myparatight{Locality loss}
Recall that  the locality goal means that the patching does not influence the feature vectors for clean inputs. In particular, we assume the provider has a clean unlabeled validation dataset $D_{val}$. For each input in $D_{val}$, the  post-patching foundation model and pre-patching one should output similar feature vectors. Therefore, our locality loss $\mathcal{L}_l$ is the average similarity between the feature vectors output by $h'$ and $h$ for the inputs in the validation dataset and the reference input. Formally, we have the following: 
\begin{align}
\label{eq:loss_locality}
\mathcal{L}_l = -\frac{1}{\left|\mathcal{D}_{\text {val }}\right|+1} \sum_{x \in\left\{\mathcal{D}_{\text {val }} \cup \{x_r\}\right\}} sim\left(h(x), h^{\prime}(x)\right),
\end{align}
where  $\left|\mathcal{D}_{\text {val }}\right|$ denotes the number of inputs in the validation dataset. We consider the reference input $x_r$ in defining the locality loss to guarantee that the feature vector of $x_r$ does not change much and it is still correctly classified by the downstream classifier. A smaller $\mathcal{L}_l$ indicates that $h$ and $h^{\prime}$ produce more similar feature vectors for a clean input.

\myparatight{Generalizability loss} Suppose we have reverse engineered a backdoor trigger $t_b$ from the bug instance ($x_b$, $x_r$) as described in Section~\ref{sec:reverse_engineer_bug_triggers}. The generalizability goal means that embedding the trigger into an input has minimal impact on its feature vector output by the post-patching foundation model. Therefore, our generalizability loss $\mathcal{L}_{g}$ is the average similarity between the feature vectors of an input and its trigger-embedded version for the reference input and inputs in the validation dataset. Formally, we have the following: 
\begin{align}
\label{eq:loss_generalizability}
\mathcal{L}_{g}=-\frac{1}{\left|\mathcal{D}_{\text {val }}\right|+1} \sum_{x \in\left\{\mathcal{D}_{\text {val }} \cup \left\{x_r\right\}\right\}} sim\left(h^{\prime}(x \oplus t_b), h^{\prime}(x)\right). 
\end{align}
A smaller $\mathcal{L}_{g}$ indicates that $h^{\prime}$ outputs more similar feature vectors for an input $x$ embedded with the reverse engineered trigger $t_b$ and its clean counterpart. 

\myparatight{Optimization problem}
After defining the three loss terms, we can now present our optimization problem, which aims to obtain a post-patching foundation model $h^{\prime}$ from a pre-patching one via minimizing a weighted sum of the three loss terms. Formally, we have the following optimization problem:
\begin{align}
\label{eq:loss_all}
\min_{h^{\prime}} \mathcal{L} = \mathcal{L}_{e} + \lambda_l \mathcal{L}_{l} +\lambda_g \mathcal{L}_{g},
\end{align}
where $\lambda_l$ and $\lambda_g$ are two hyperparameters that balance the three loss terms. By minimizing this combined loss function $\mathcal{L}$, our \method finds a post-patching foundation model $h^{\prime}$ that achieves the three patching goals simultaneously.

\subsection{Solving the Optimization Problem}
Patching a foundation model $h$ is to solve the optimization problem in Equation~\ref{eq:loss_all}, which turns $h$ into a post-patching  foundation model $h^{\prime}$. We propose a Stochastic Gradient Descent based algorithm to solve the optimization problem, as shown in Algorithm~\ref{alg:ours}. Specifically, we initialize $h^{\prime}$ to be $h$. Then, we iteratively update $h^{\prime}$ using the gradient of the loss function $\mathcal{L}$ with respect to a mini-batch of the validation dataset. The iterative process is repeated for $T$ epochs. 

\begin{algorithm}[!t]
\caption{Our \method}\label{alg:ours}
\begin{algorithmic}[1]
\Require Bug instance $(x_b,x_r)$,  pre-patching foundation model $h$, validation dataset $\mathcal{D}_{\text{val}}$, learning rate $\alpha$, mini-batch size $m$, number of epochs $T$,  and reverse engineered trigger  $t_b$
\Ensure Post-patching foundation model $h^{\prime}$
\State Initialize $h^{\prime} \leftarrow h$
\State $I = \lceil \frac{|\mathcal{D}_{\text{val}} \cup \{x_r\}|}{m} \rceil$ \Comment{Number of iterations per epoch}
\For{$t = 1, 2, \dots, T$}
\For{$i = 1, 2, \dots, I$}
\State Sample a mini-batch $\mathcal{B} \subset \mathcal{D}_{val} \cup \{x_r\}$ s.t. $|\mathcal{B}|=m$
\State Compute gradient $\nabla_{h^{\prime}} \mathcal{L}$ 
\State $h^{\prime} \leftarrow h^{\prime} - \alpha \nabla_{h^{\prime}} \mathcal{L}$
\EndFor
\EndFor
\State \Return $h^{\prime}$
\end{algorithmic}
\end{algorithm}

\subsection{Reverse Engineering a Trigger}
\label{sec:reverse_engineer_bug_triggers}
The generalizability loss requires the backdoor trigger in the misclassified input $x_b$. However, a bug instance only consists of a pair of inputs.  To address the challenge, we propose a method to automatically reverse engineer the backdoor trigger $t_b$ from a bug instance, where the trigger is a set of pixels in an image input or a set of words at particular locations in a text input. Our key observation is that the backdoor trigger is the major cause for the dissimilarity between the feature vectors of $x_b$ and $x_r$ and thus the misclassification of $x_b$.  Based on this observation, we leverage an interpretable machine learning method to calculate an \emph{attribution score} for each pixel in an image input or each word in a text input, where an attribution score aims to quantify the influence of a pixel/word on the dissimilarity between the feature vectors of $x_b$ and $x_r$. Intuitively, the trigger pixels/words have large attribution scores. Therefore, we further use a clustering algorithm to identify the pixels/words with large attribution scores and treat them  as a trigger. Next, we describe how to calculate attribution scores and identify a trigger based on them. 

\myparatight{Step I: calculating attribution scores}
Suppose we are given a  pre-patching foundation model $h$ and a bug instance $(x_b, x_r)$. $x_b$ is misclassified by the downstream classifier because of the trigger in it. Specifically, the trigger pixels/words in $x_b$ contribute substantially to the dissimilarity between the feature vectors $h(x_b)$ and \hongbin{$h(x_r)$}, and thus the misclassification of $x_b$. This observation inspires us to leverage an interpretable machine learning method~\cite{zeiler2014visualizing,selvaraju2017grad,simonyan2013deep,springenberg2014striving} to quantify the influence of a pixel/word in  $x_b$. Given a machine learning model, an input $x_b$, and an \emph{objective function} about the input, an interpretation method can calculate an attribution score for each pixel/word in the input, where the attribution score quantifies the contribution of a pixel/word on the objective function. For instance, the \emph{occlusion} method~\cite{zeiler2014visualizing}  systematically occludes portions of the input, such as rectangular regions of an image input or text sequences in a text input. By doing so and calculating the difference in the objective function, the occlusion method can calculate an attribution score for each pixel/word.  

\begin{algorithm}[!t]
\caption{Reverse Engineering a Trigger}
\label{alg:reverse_engineer_bug_trigger}
\begin{algorithmic}[1]
\Require
Bug instance ($x_b$, $x_r$),
pre-patching foundation model $h$, 
objective function $\ell$,
and interpretation method $\mathcal{A}$
\Ensure Trigger $t_b$.

\State \textbf{Step I: Calculate Attribution Scores}
\State $\ell_b \gets \ell(h,x_b, \hongbin{x_r})$ \Comment{As defined in Equation~\ref{attribution_obj} }
\State $A \gets \mathcal{A}(h,\ell_b,x_b)$ \Comment{Attribution scores}
\State \textbf{Step II: Identify  Trigger}
\State $K \gets 2$ \Comment{Number of clusters}
\State $C_1,\  C_2 \gets \text{K-means}(A, K)$ 
\State $M_1,\  M_2\gets \text{Mean attribution score of each cluster}$
\State $i \gets \arg\max_{j} M_j$ \Comment{Find the higher mean score}
\State $t_b \gets C_i$ \Comment{Reverse engineered trigger}
\State \Return $t_b$
\end{algorithmic}
\end{algorithm}

\begin{figure}[!t]
  \centering
  \includegraphics[width=0.45\textwidth]{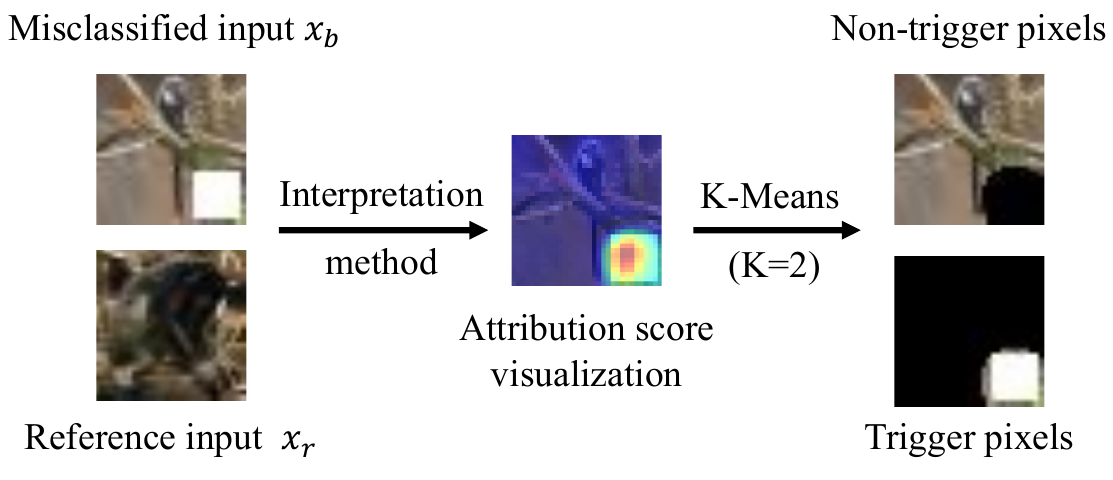}
  \caption{Illustration of reverse engineering a trigger.}
  \label{fig:validation-size}
\end{figure}

A key challenge of applying interpretation methods to reverse engineer a trigger is how to design the objective function. One straightforward objective function is the loss function of the  label $f \bigodot h(x_b)$ predicted by the downstream classifier for the input $x_b$. However, the foundation-model provider does not have access to the downstream classifier $f$. As a result, the provider cannot evaluate the loss function of $f \bigodot h(x_b)$ nor apply an interpretation method to calculate the attribution scores. To address the challenge, we define the objective function using the foundation model and bug instance  without involving the downstream classifier. In particular, our objective function is directly related to the dissimilarity between the feature vectors $h(x_b)$ and {$h(\hongbin{x_r})$. Formally, we define the objective function $\ell(h,x_b, \hongbin{x_r})$ as follows: 
\begin{align}
\label{attribution_obj}
    \ell(h,x_b, \hongbin{x_r}) = 1-sim(h(x_b), h(\hongbin{x_r})),
\end{align}
where $sim$ is the cosine similarity in our experiments. Given such objective function, we leverage an interpretation method to calculate an attribution score for each pixel/word in $x_b$. A higher attribution score is likely to indicate that the corresponding pixel/word has a larger impact on the objective function, i.e., the dissimilarity between $h(x_b)$ and $h(\hongbin{x_r})$. 

We note that several previous studies (e.g.,~\cite{doan2020februus,chou2020sentinet}) also leverage interpretable machine learning methods to identify the backdoor trigger in an input. However, they focus on classifiers (i.e., $f \bigodot h$ in our problem) and use the straightforward loss function of the predicted label as the objective function. Our work focuses on foundation models and designs a new objective function tailored to them.  

\myparatight{Step II: identifying trigger} 
After we have obtained the attribution score for each input pixel/word, we identify the trigger pixels/words based on them. Intuitively, the input pixels/words with larger attribution scores may form the  trigger. Based on this intuition, we use a clustering method to divide the pixels/words into two clusters: one cluster containing the pixels/words with high attribution scores, and one cluster containing those with low attribution scores. For instance, we can leverage the K-means algorithm~\cite{macqueen1967classification} with $K=2$ to perform the clustering. 
We then treat the cluster with the higher mean attribution score as the trigger. We summarize our method to reverse engineer a trigger in Algorithm~\ref{alg:reverse_engineer_bug_trigger}.

\section{Evaluation}
\label{sec:evaluation}

\subsection{Experimental Setup}

\myparatight{Datasets}
We use eleven benchmark datasets, including six image datasets, two image-text datasets, and three text datasets. We show the details of these datasets in Table~\ref{tab:data_statistics} in Appendix. These benchmark datasets were also used in prior studies~\cite{jia2022badencoder,zhang2022corruptencoder,carlinipoisoning,shen2021backdoor,zhang2023red} on backdoor attacks to foundation models. In addition to the numbers shown in Table~\ref{tab:data_statistics} in Appendix, the STL10 dataset 
also contains 100,000 unlabeled images. We use these unlabeled images for pre-training when STL10 is used as the pre-training dataset. We follow Zhang et al.~\cite{zhang2022corruptencoder} to construct ImageNet100-A and ImageNet100-B, two subsets of ImageNet. Each of these subsets contains randomly sampled 100 classes from the ImageNet dataset with no overlapping classes.  The CLIP-400M dataset was used for pre-training the OpenAI's CLIP foundation model, which contains 400M image-text pairs but is not publicly available. The CC3M-Sub and Wiki103-Sub denote randomly subsampled subsets of the CC3M~\cite{sharma2018conceptual} and Wiki103~\cite{merity2016pointer}, respectively.

\myparatight{Backdoor attacks to foundation models}
We consider the following five state-of-the-art backdoor attacks to foundation models. 
    \textbf{\textit{BadEncoder~\cite{jia2022badencoder}.}} BadEncoder injects backdoors into a vision foundation model by directly modifying its model parameters. A backdoored vision foundation model outputs  attacker-desired feature vectors for any inputs embedded with a trigger. These feature vectors then cause a downstream classifier to produce attacker-desired predictions. BadEncoder is applicable to both single-modal and multi-modal vision foundation models.
    \textbf{\textit{CorruptEncoder~\cite{zhang2022corruptencoder}.}} CorruptEncoder injects backdoors into a vision foundation model by poisoning its pre-training data. In particular, CorruptEncoder injects a small fraction of poisoning inputs into the pre-training dataset such that the learnt vision foundation model outputs attacker-desired feature vectors for any inputs embedded with a trigger.  Like BadEncoder, CorruptEncoder is also applicable to both single-modal and multi-modal vision foundation models.

    \textbf{\textit{Carlini \& Terzis~\cite{carlinipoisoning}.}}  Carlini and Terzis proposed a  backdoor attack to multi-modal vision foundation models. Their attack first creates a collection of text captions that include the target class name chosen by the attacker. Then their attack associates these text captions with a set of images embedded with the trigger to construct  poisoning image-text pairs, which are then injected into the pre-training dataset. 

    \textbf{\textit{POR~\cite{shen2021backdoor} and NeuBA~\cite{zhang2023red}.}} These are two backdoor attacks to language foundation models. POR  modifies a  language foundation model's parameters to output a specific feature vector for any input containing a trigger, while NeuBA introduces a new backdoor pre-training objective in addition to the original pre-training objective to achieve the same  goal.

\myparatight{Pre-training (backdoored) foundation models}
We pre-train (backdoored) foundation models  following the default experimental settings of the aforementioned backdoor attacks in the original papers, or use the publicly available ones from their codebase. For POR and NeuBA, we adopt a unified implementation~\cite{cui2022unified}. Table~\ref{tab:pre_training_settings} in Appendix summarizes the pre-training settings and backdoor triggers. 

\myparatight{Training downstream classifiers}
We use a foundation model as a feature extractor to train downstream classifiers on downstream datasets. For each downstream dataset, we use the training examples to train the downstream classifier and use the testing examples to evaluate its performance. These downstream classifiers are fully-connected neural networks trained with cross-entropy loss and the Adam optimizer. We follow the default parameter settings as those in the original papers on backdoor attacks when training the downstream classifiers. Table~\ref{tab:downstream_training_settings} in Appendix summarizes the training settings of downstream classifiers.

\myparatight{Compared patching methods}
We compare \method with the following patching methods. 

    \textbf{\textit{Fine-tuning (FT).}} FT is a widely used method for slightly modifying a model. 
    When patching a foundation model, FT can achieve the effectiveness goal by optimizing the foundation model's parameters to produce similar feature vectors for $x_b$ and $x_r$. Specifically, FT obtains the post-patching foundation model $h^{\prime}$ by solving the optimization problem: $\min_{h^{\prime}}  \mathcal{L}_e$, where $\mathcal{L}_e$ is our effectiveness loss defined in Equation~\ref{eq:effectiveness_loss}. $h^{\prime}$ is initialized as $h$ and solved iteratively by gradient descent. We use the same parameter setting for FT as our~\methodtight.

    \textbf{\textit{Fine-tuning with $\ell_2$-norm or $\ell_{\infty}$-norm constraint (FT+$\ell_2$ or FT+$\ell_{\infty}$).}} Zhu et al.~\cite{zhu2020modifying} use FT with an $\ell_2$-norm or $\ell_{\infty}$-norm constraint to address overfitting and catastrophic forgetting. We apply this method to patch foundation models. 
    In particular, when patching a foundation model $h$, FT+$\ell_{p}$ obtains $h^{\prime}$ by solving the following optimization problem: $\min_{h^{\prime}}  \mathcal{L}_e \text{ subject to } \ell_{p}(h^{\prime},h) \leq \delta$, 
    where $p=2$ or $\infty$ and $\delta$ is the threshold for the $\ell_2$ or $\ell_{\infty}$ constraint. We use projected gradient descent to solve the optimization problem. Specifically, during each iteration, we first compute the gradient of the loss function with respect to the model parameters; after updating the model parameters, we project them onto the feasible set defined by the $\ell_p$ constraint if needed. FT+$\ell_2$ and FT+$\ell_{\infty}$ use the same parameter settings as FT. We set a small threshold $\delta=0.01$ to bound the model parameters' change. 

    \hongbin{\textbf{\textit{Unlearning.}} Liu et al.~\cite{liu2022backdoor} propose a machine unlearning based pre-deployment defense to erase backdoors from classifiers. They fine-tune a backdoored classifier to maximize the cross-entropy loss on trigger-embedded inputs with the attacker's target label. Furthermore, they bound the changes in model parameters using an $\ell_1$ norm constraint. We extend their method to patch foundation models by replacing the cross-entropy loss as a cosine similarity loss  between the reference input $x_r$ and misclassified input $x_b$.} 
    
    \hongbin{\textbf{\textit{Fine-Pruning.}} Liu et al.~\cite{liu2018fine} combine pruning and fine-tuning as a pre-deployment defense to eliminate backdoors from classifiers. This defense first prunes the  channels in the convolutional layers that are dormant, i.e., that have negative average activation on  clean  data; and then it fine-tunes the pruned model to regain utility lost from pruning. We extend fine-pruning to patch foundation models as follows: we first prune 50\% of the channels that are dormant on the reference input $x_r$ and then fine-tune the pruned foundation model using the pre-training data for 50 epochs.}

\begin{table*}[!t]
  \centering
  \vspace{3mm}
 \fontsize{7.5}{10}\selectfont
   \caption{Results for patching vision foundation models.}
	\label{patch_single_modal}
	\subfloat[Single-modal]{
    \begin{tabular}{|c|c|c|c|c|c|c|c|c|c|c|}
    \hline
    \multirow{3}{*}{\makecell{Attack method}}  & \multirow{3}{*}{\makecell{Pre-training\\ dataset}} & \multirow{3}{*}{\makecell{Downstream\\dataset}} & \multicolumn{4}{c|}{{\makecell{Before patching}}}    & \multicolumn{4}{c|}{\makecell{After patching}}  \\
    \cline{4-11}   &    &   &   \multirow{2}{*}{{\makecell{CP}}}  & \multicolumn{1}{c|}{\multirow{2}{*}{Acc} }   &  \multicolumn{1}{c|}{\multirow{2}{*}{ASR } } & \multicolumn{1}{c|}{\multirow{2}{*}{AccB } }  &    \multirow{2}{*}{{\makecell{CP}}} &  \multicolumn{1}{c|}{\multirow{2}{*}{Acc } }  &  \multicolumn{1}{c|}{\multirow{2}{*}{ASR } } & \multicolumn{1}{c|}{\multirow{2}{*}{AccB } } \\
    & & & & & & & & & & \\
    \hline  
    \hline
    \multirow{4}{*}{\makecell{BadEncoder}}   & \multirow{2}{*}{CIFAR10} &  STL10    & $\times$     & 76.46  & 99.82      & 0.10   &  \checkmark   &76.59    & 2.39    &   73.38  \\
    \cline{3-11}
        &   & SVHN  & $\times$  & 69.07   & 98.92      &0.53      & \checkmark   &   78.49    & 5.97    &   65.39   \\
    \cline{2-11}
    & \multirow{2}{*}{STL10} &  CIFAR10 & $\times$       & 86.64    & 97.07      &1.24   & \checkmark   &86.39    & 2.59    &   81.20    \\
    \cline{3-11}
        &   & SVHN  & $\times$   & 65.21   & 97.44      &0.36  & \checkmark    &76.21    & 7.94    &   61.04    \\
    \hline  \hline
    \multirow{2}{*}{CorruptEncoder}  & \multirow{2}{*}{ImageNet100-A} 
    &  {ImageNet100-B}   & $\times$    & 61.68     & 94.32      &2.57    &  \checkmark   &61.76    &1.57    & 57.64    \\
    \cline{3-11}
        &   & Oxford-IIIT Pets  & $\times$   &  56.71   &   71.90    &  4.96 &   \checkmark   &  55.69    &  2.18  &  51.23 \\
    \hline
    \end{tabular}
    }

	\subfloat[Multi-modal]{ \label{patch_multi_modal}
    \begin{tabular}{|c|c|c|c|c|c|c|c|c|c|c|}
    \hline
    \multirow{3}{*}{\makecell{Attack method}}  & \multirow{3}{*}{\makecell{Pre-training\\ dataset}} & \multirow{3}{*}{\makecell{Downstream\\dataset}} & \multicolumn{4}{c|}{{\makecell{Before patching}}}    & \multicolumn{4}{c|}{\makecell{After patching}}  \\
    \cline{4-11}   &    &   &   \multirow{2}{*}{{\makecell{CP}}}  & \multicolumn{1}{c|}{\multirow{2}{*}{Acc} }   &  \multicolumn{1}{c|}{\multirow{2}{*}{ASR } } & \multicolumn{1}{c|}{\multirow{2}{*}{AccB } }  &    \multirow{2}{*}{{\makecell{CP}}} &  \multicolumn{1}{c|}{\multirow{2}{*}{Acc } }  &  \multicolumn{1}{c|}{\multirow{2}{*}{ASR } } & \multicolumn{1}{c|}{\multirow{2}{*}{AccB } } \\
    & & & & & & & & & & \\
    \hline
    \hline
   \multirow{2}{*}{\makecell{BadEncoder}}   & \multirow{2}{*}{CLIP-400M} &  STL10     & $\times$    & 96.70   &99.92     & 0.04   &   \checkmark     & 95.65   & 0.42    & 95.06     \\
    \cline{3-11}
            &   & SVHN  &   $\times$  & 69.93     & 100.00      & 0.00      & \checkmark    &73.35     & 3.48   & 69.13  \\
    \hline  \hline
    \multirow{1}{*}{\makecell{Carlini \& Terzis}}   & \multirow{1}{*}{CC3M-Sub} &  ImageNet100-B     &  $\times$    &  51.47     &   98.22   &  1.09   & \checkmark   &  49.26  & 5.03  & 41.53  \\
    \hline  \hline
    \multirow{1}{*}{CorruptEncoder}  & \multirow{1}{*}{CC3M-Sub} &  ImageNet100-B     &   $\times$   &  48.44     &  94.40    &    0.81 &  \checkmark  & 45.92 & 3.46 & 40.27   \\
    \hline
    \end{tabular}
    }
\end{table*}

\begin{table*}[!t]
  \centering
  \fontsize{7.5}{10}\selectfont
  \vspace{3mm}
   \caption{Results for patching langugae foundation models.}
	\label{patch_LM}
	{
    \begin{tabular}{|c|c|c|c|c|c|c|c|c|c|c|}
    \hline
    \multirow{3}{*}{\makecell{Attack method}}  & \multirow{3}{*}{\makecell{Pre-training\\ dataset}} & \multirow{3}{*}{\makecell{Downstream\\dataset}} & \multicolumn{4}{c|}{{\makecell{Before patching}}}    & \multicolumn{4}{c|}{\makecell{After patching}}  \\
    \cline{4-11}   &    &   &   \multirow{2}{*}{{\makecell{CP}}}  & \multicolumn{1}{c|}{\multirow{2}{*}{Acc} }    &  \multicolumn{1}{c|}{\multirow{2}{*}{ASR } } & \multicolumn{1}{c|}{\multirow{2}{*}{AccB } }  &    \multirow{2}{*}{{\makecell{CP}}} &  \multicolumn{1}{c|}{\multirow{2}{*}{Acc } }  &  \multicolumn{1}{c|}{\multirow{2}{*}{ASR } } & \multicolumn{1}{c|}{\multirow{2}{*}{AccB } } \\
    & & & & & & & & & & \\
    \hline
    \hline 
    \multirow{2}{*}{\makecell{NeuBA}}   & \multirow{2}{*}{Wiki-103} &   SST-2    &   $\times$   &   80.28    &  100.00    &  0.00   &  \checkmark  &  79.68	&	24.34	& 	75.65 \\\cline{3-11}
    &   &   HSOL     & $\times$  & 80.16  &  100.00 &  0.00 & \checkmark  & 81.28	&	18.28	& 	81.72 \\
    \hline 
    \hline 
    \multirow{2}{*}{\makecell{POR}}   & \multirow{2}{*}{Wiki-103} &   SST-2    &  $\times$    &     82.59     &   67.66   &   12.34  &  \checkmark  &  81.82	&	17.49	& 	82.51 \\\cline{3-11}
    &   &   HSOL    & $\times$   & 82.90  & 99.92  & 0.08  &  \checkmark   &  82.78	&	17.07	& 	82.93 \\\hline

    \end{tabular}
    }
      \vspace{-1.5mm}
\end{table*}

\myparatight{Evaluation metrics} We use four evaluation metrics, i.e., Correct Prediction of $x_b$ (CP), Accuracy (Acc), Attack Success Rate (ASR), and Accuracy with Backdoor trigger (AccB). CP evaluates the effectiveness goal, Acc evaluates the locality goal, ASR and AccB evaluate the generalizability goal. 
We define the four evaluation metrics as follows:

    \textbf{\textit{Correct Prediction of $x_b$ (CP).}} CP measures whether  a downstream classifier correctly classifies  $x_b$.  CP can be  either $\times$ or \checkmark, denoting incorrect or correct classification of $x_b$.  
    
    \textbf{\textit{Accuracy (Acc).}} Acc denotes the testing accuracy of a downstream classifier $f\bigodot h$ (or $f' \bigodot h'$ after patching). 
    Specifically, Acc is the fraction of clean downstream testing examples that are correctly classified. If post-patching Acc is  close to or higher than pre-patching Acc, then the locality goal is achieved. 
    
    \textbf{\textit{Attack Success Rate (ASR).}} 
    We generate \emph{backdoored testing inputs} by embedding a trigger into all clean downstream testing inputs that are not from the target class.  ASR is the fraction of such backdoored testing inputs that are classified as the target class by the downstream classifier. 
    
    \textbf{\textit{Accuracy with backdoor trigger (AccB).}} AccB is the testing accuracy of a downstream classifier $f\bigodot h$ (or $f' \bigodot h'$ after patching) 
    for backdoored testing inputs. In particular, AccB is the fraction of backdoored testing inputs that are correctly classified by a downstream classifier.  
    The generalizability goal is achieved if the post-patching ASR is small and AccB is close to Acc.

\myparatight{Parameter settings of \methodtight}  We randomly sample a bug instance $(x_b, x_r)$, where $x_b$ is a trigger-embedded misclassified testing input and $x_r$ is a correctly classified testing input  from the downstream testing dataset. By default, we use BadEncoder as the backdoor attack,  CIFAR10 for pre-training, and STL10 for downstream classifier training.

The default parameter settings for our \method are as follows: $\lambda_l=\lambda_g=1$; the number of epochs $T=200$ in  Algorithm~\ref{alg:ours}; and the learning rates used for patching are shown in Table~\ref{tab:patching_settings} in Appendix. The  validation dataset size  and batch size are determined based on our computation resources. Table~\ref{tab:patching_settings} in Appendix summarizes these parameters. Unless otherwise mentioned, we sample the validation dataset from the pre-training dataset, except for CLIP-400M. Since CLIP-400M is not publicly available, we sample the training inputs in ImageNet as our validation dataset. By default, we use the Occlusion interpretation method~\cite{zeiler2014visualizing} when reverse engineering a trigger. \hongbin{We performed experiments using 18 NVIDIA RTX 6000 GPUs, with each GPU having 24GB memory.}

\subsection{Experimental Results}

\myparatight{\method achieves the three patching goals}   Table~\ref{patch_single_modal} and Table~\ref{patch_LM} show the results of patching vision and language foundation models, respectively. We have multiple observations. First, \method consistently achieves the effectiveness goal in all settings, indicated by CP~=~\checkmark. This is because the post-patching foundation model produces similar feature vectors for the misclassified input $x_b$ and reference input $x_r$. 

Second, \method also achieves the locality goal by maintaining  Acc after patching. This is because, for a clean input,  the post-patching and pre-patching foundation models output similar feature vectors. In particular, even when the validation dataset is a  small fraction of the clean pre-training dataset, it is sufficient for our~\method to achieve the locality goal. For example, as shown in Table~\ref{patch_multi_modal}, when the pre-training dataset is CLIP-400M and the foundation model is attacked by BadEncoder, using only 50,000 images (only 0.0125\% of 400 million) from ImageNet as the validation dataset is sufficient to achieve the locality goal.

Third, our~\method successfully achieves the generalizability goal since ASR is low (close to 0) {except for patching language models} and AccB is close to Acc after patching  across all settings. {We note that ASR is not close to 0 for patching language foundation models in Table~\ref{patch_LM}. This is because the downstream datasets have two classes. Given a downstream classifier built based on a clean language foundation model,  ASR  would be close to the misclassification rate of testing inputs. We observe that ASR in Table~\ref{patch_LM} is indeed close to the misclassification rate of testing inputs, roughly 20\%. Therefore, the post-patching language foundation models are similar to clean language foundation models. Our results suggest that our method effectively mitigates the backdoor vulnerabilities in a backdoored foundation model. }

\myparatight{\method outperforms existing methods} 
Table~\ref{tab:baseline_patch_single_modal} compares \method with existing patching methods for patching the vision foundation model in our default setting. We observe that \method  outperforms these methods at achieving the three patching goals. 
\hongbin{Specifically, FT, FT+$\ell_2$, FT+$\ell_\infty$, and unlearning can achieve the effectiveness goal. This is because they all fine-tune the foundation model to minimize the effectiveness loss $\mathcal{L}_e$. They achieve the locality goal to some extent, i.e., Acc is close to that before patching, but they cannot achieve the generalizability goal. Fine-Pruning achieves the locality and generalizability goals to some extent, but it cannot achieve the effectiveness goal.  }

\begin{table}[!t]
  \centering
  \vspace{3mm}
  \fontsize{8}{10}\selectfont
   \caption{Comparing \method with existing methods.}
	\label{tab:baseline_patch_single_modal}
	{
    \begin{tabular}{|c|c|c|c|c|}
     \hline
    \multirow{2}{*}{\makecell{Patching method}}   & \multirow{2}{*}{\makecell{CP}}   & \multirow{2}{*}{\makecell{Acc}}  &  \multirow{2}{*}{\makecell{ASR }}    & \multirow{2}{*}{\makecell{AccB}}     \\
    & & & & \\
    \hline  \hline
    {\makecell{Before patching}}  &  $\times$     & 76.46    & 99.82      & 0.10     \\  \hline
    {\makecell{FT}}  &     \checkmark     &  69.61  & 40.57    &   32.50  \\  \hline
    {\makecell{FT+$\ell_2$}}  &   \checkmark     &    72.05 & 55.90    &   24.31   \\  \hline
    {\makecell{FT+$\ell_\infty$}}  &   \checkmark     &   70.04  & 41.79    &   30.61  \\  \hline
    {\makecell{\hongbin{Unlearning}}}  &    \hongbin{\checkmark}     &   \hongbin{70.02} &  \hongbin{34.57}   &   \hongbin{32.12}  \\  \hline
    {\makecell{\hongbin{Fine-Pruning}}}  &    \hongbin{$\times$}     &  \hongbin{71.92}    & \hongbin{6.65}    &   \hongbin{61.97}    \\  \hline
    {\makecell{\methodtight}}  &  \checkmark   &76.59    & 2.39    &   73.38  \\  \hline
    \end{tabular}%
    }
\end{table}

\myparatight{Impact of different interpretation methods}
Our~\method uses an interpretation method to reverse engineer a trigger  from a bug instance. We compare  four widely used interpretation methods, i.e.,  Occlusion~\cite{zeiler2014visualizing}, GradCam~\cite{selvaraju2017grad}, Saliency Map~\cite{simonyan2013deep}, Guided Backprobagation (GuidedBack)~\cite{springenberg2014striving}. Table~\ref{impact_interpret_ml} shows the results, from which we have the following two main observations. First, we observe that different interpretation methods lead to different reverse-engineered  triggers. Specifically, Occlusion and GradCam can reverse-engineer triggers that contain majority of the true backdoor trigger  (the white square at the bottom right corner). On the other hand, the triggers reverse-engineered by Saliency Map and GuidedBack contain only a part of the true backdoor trigger. Therefore, Occlusion and GradCam outperform the other two in terms of the generalizability goal, i.e.,  ASRs are lower and AccBs are higher. Second, we find that \method successfully achieves the effectiveness and locality goals, regardless of the interpretation method used. The reason is that the reverse-engineered trigger is used only in the generalizability loss and does not impact the effectiveness and locality losses.

\begin{table}[!t]
  \centering
  \fontsize{7.5}{10}\selectfont
  \vspace{5mm}
  \caption{Impact of different interpretation methods.}
	\label{impact_interpret_ml}
	{
    \begin{tabular}{|c|c|c|c|c|c|}
     \hline
    \multirow{2}{*}{\makecell{Interpretation\\method}}    &  \multirow{2}{*}{\makecell{Reverse engineered\\trigger}}  & \multirow{2}{*}{{\makecell{CP}}}   & \multirow{2}{*}{\makecell{Acc}} &  \multirow{2}{*}{\makecell{ASR }}    & \multirow{2}{*}{\makecell{AccB}}     \\
    & & & & &\\
    \hline  \hline
    \multirow{2}{*}{\makecell{Occlusion}}  & \multirow{2}{*}{\makecell{\adjustbox{valign=c}{\includegraphics[width=0.03\textwidth]{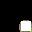}}}} & \multirow{2}{*}{\checkmark}    & \multirow{2}{*}{76.59}    &  \multirow{2}{*}{2.39}     &    \multirow{2}{*}{73.38}       \\
    &   &   &   &   &   \\\hline
    \multirow{2}{*}{\makecell{GradCam}}  & \multirow{2}{*}{\makecell{\adjustbox{valign=c}{\includegraphics[width=0.03\textwidth]{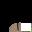}}}} & \multirow{2}{*}{\checkmark}    & \multirow{2}{*}{76.47}    &  \multirow{2}{*}{3.42}     &    \multirow{2}{*}{71.44}       \\
    &   &   &   &   &   \\\hline
    \multirow{2}{*}{\makecell{Saliency Map}}  & \multirow{2}{*}{\makecell{\adjustbox{valign=c}{\includegraphics[width=0.03\textwidth]{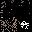}}}} & 
    \multirow{2}{*}{\checkmark}    & \multirow{2}{*}{76.58}    &  \multirow{2}{*}{11.88}     &    \multirow{2}{*}{61.72}       \\
    &   &   &   &   &   \\\hline
    \multirow{2}{*}{\makecell{GuidedBack}}  & \multirow{2}{*}{\makecell{\adjustbox{valign=c}{\includegraphics[width=0.03\textwidth]{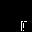}}}} & 
    \multirow{2}{*}{\checkmark}    & \multirow{2}{*}{76.29}    &  \multirow{2}{*}{13.56}     &    \multirow{2}{*}{61.67}       \\
    &   &   &   &   &   \\\hline
    \end{tabular}%
    }
\end{table}

\begin{table}[!t]
  \centering
  \vspace{3mm}
  \fontsize{8}{10}\selectfont
  \caption{Impact of the three loss terms.}
	\label{impact_loss_term}
	{
    \begin{tabular}{|c|c|c|c|c|}
     \hline
    \multirow{2}{*}{\makecell{Removed loss term}}   & \multirow{2}{*}{\makecell{CP}}   & \multirow{2}{*}{\makecell{Acc}} &  \multirow{2}{*}{\makecell{ASR}}    & \multirow{2}{*}{\makecell{AccB}}     \\
    & & & &\\
    \hline  \hline
       None &  \checkmark   & 76.59     &2.39    & 73.38     \\  \hline
    {\makecell{Effectiveness loss}}  &  $\times$     &  76.70    & 2.65    &   72.97      \\  \hline
 {\makecell{Locality loss}}  &  \checkmark   &   48.50    & 7.10    &   45.61     \\  \hline
   {\makecell{Generalizability loss}}    &   \checkmark   &  76.70    & 26.49    &   50.57       \\  \hline
    \end{tabular}%
    }
\end{table}

\myparatight{Impact of the three loss terms}
Our formulated optimization problem in Equation~\ref{eq:loss_all} consists of three loss terms. We study the impact of removing each of them and show the results in Table~\ref{impact_loss_term}. We have the following three key observations. First, the removal of the effectiveness loss hinders the patched foundation model's ability to correctly classify the misclassified input $x_b$. Second, removing the locality loss leads to a substantial drop in Acc, falling from 76.59\% (with no loss term removed) to 48.50\%. This highlights the critical role of the locality loss in preserving the utility of the patched foundation models. Third, the removal of the generalizability loss results in a significant increase in ASR to 26.49\% and a decrease in AccB to 50.57\%. This indicates the significant role of the generalizability loss in mitigating backdoor vulnerabilities. To summarize, these observations emphasize the importance of integrating all the three loss terms to achieve the three patching goals. 

\begin{figure}[!t]
  \centering
  \subfloat[STL10]{\includegraphics[width=0.24\textwidth]{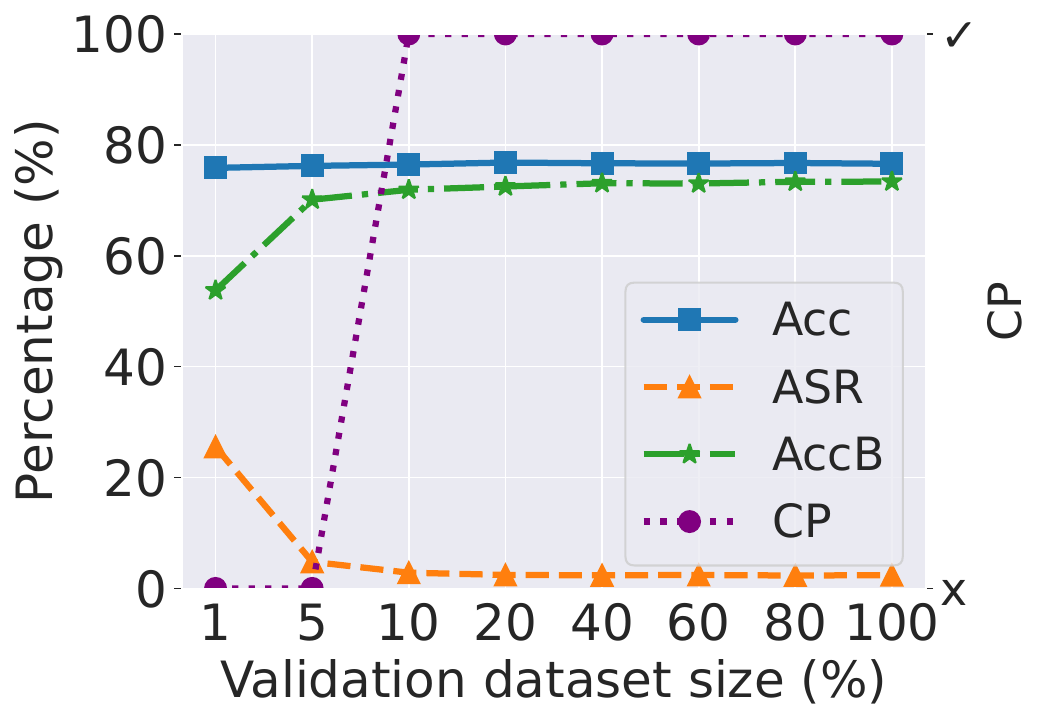}\label{fig:validation-size-stl10}}
  \subfloat[SVHN]{\includegraphics[width=0.24\textwidth]{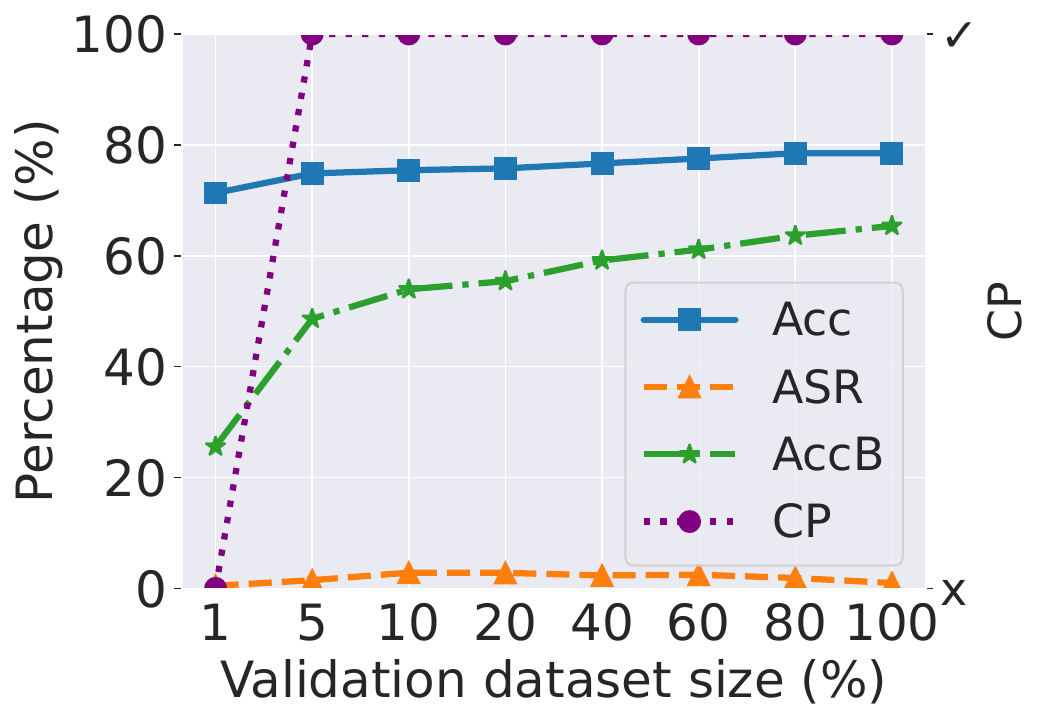}\label{fig:validation-size-svhn}}
  \caption{Impact of the validation dataset size on different downstream datasets. The validation dataset is a subset of the  pre-training dataset CIFAR10.}
  \label{fig:validation-size}
\end{figure}

\myparatight{Impact of validation dataset size}
\method relies on a validation dataset $\mathcal{D}_{\text{val}}$ to calculate both the locality loss (Equation~\ref{eq:loss_locality}) and the generalizability loss (Equation~\ref{eq:loss_generalizability}). We investigate  how varying the size of the validation dataset influences our method's performance on three downstream datasets, namely STL10 and SVHN. We show the results in Figure~\ref{fig:validation-size}. We find that both CP and Acc converge when  the size of the validation dataset exceeds 10\% of the pre-training dataset size across the tested downstream datasets. This suggests that our method is efficient in achieving its effectiveness and locality goals, even when the validation dataset is a small fraction of the pre-training dataset. Interestingly, in terms of ASR and AccB, we notice some variations. For the STL10 dataset, both metrics begin to stabilize when the validation dataset size exceeds 10\% of the pre-training dataset size. However, for the SVHN dataset, ASR stabilizes when the validation dataset is only 1\% of the pre-training dataset, while AccB continues to improve as the validation dataset size increases. This is probably because STL10's data distribution is more similar to the pre-training dataset CIFAR10, compared to SVHN.

\myparatight{Impact of validation dataset distribution} 
We further investigate the impact of different distributions of the validation dataset on our method's performance. We show the results in Table~\ref{tab:impact_valid_data_distribution}. We observe that \method achieves all three patching goals regardless of the validation dataset distribution. The reason is that all these  validation datasets contain diverse images, which  help our method achieve the locality and generalizability goals that rely on $\mathcal{D}_{val}$.

\begin{table}[!t]
  \centering
  \fontsize{8}{10}\selectfont
  \vspace{5mm}
  \caption{Impact of the validation dataset distribution. The pre-training dataset is CIFAR10.}
	\label{tab:impact_valid_data_distribution}
	{
    \begin{tabular}{|c|c|c|c|c|}
     \hline
    \multirow{2}{*}{\makecell{Validation dataset distribution}}   & \multirow{2}{*}{{\makecell{CP}}}   & \multirow{2}{*}{\makecell{Acc}} &  \multirow{2}{*}{\makecell{ASR }}    & \multirow{2}{*}{\makecell{AccB}}     \\
    & & & &\\
    \hline  \hline
 CIFAR10  & \checkmark &   76.59     & 2.39     & 73.38       \\  \hline
 STL10  & \checkmark &   76.60    & 2.71    &   72.82       \\  \hline
 ImageNet   & \checkmark    & 75.88    & 2.44    &   72.24   \\ \hline
    \end{tabular}%
    }
\end{table}

\myparatight{Patching multiple bugs}
A client may detect and report bugs to the foundation-model provider at different times. Therefore, the foundation-model provider may apply our method to  patch the foundation model multiple times. We illustrate this scenario in Figure~\ref{fig:patch_sequential}. In our experiments, the foundation-model provider receives three bug instances sequentially and patches the pre-patching foundation model $h$ three times, which results in three post-patching foundation models $h^{\prime}_1$, $h^{\prime}_2$, and  $h^{\prime}_3$.  Table~\ref{tab:impact_patch_sequential}  shows our method's performance in this case. We have two main observations. First, \method can achieve the effectiveness and locality goals in every patching attempt.  Second, patching  multiple bugs can better achieve  the generalizability goal as ASR decreases when patching more times. This is because as the foundation model is patched more times,  the trigger is better reverse-engineered (as shown in Figure~\ref{fig:patch_sequential}), which improves the generalizability of \methodtight.

\begin{figure}[!t]
  \centering
  \includegraphics[width=0.5\textwidth]{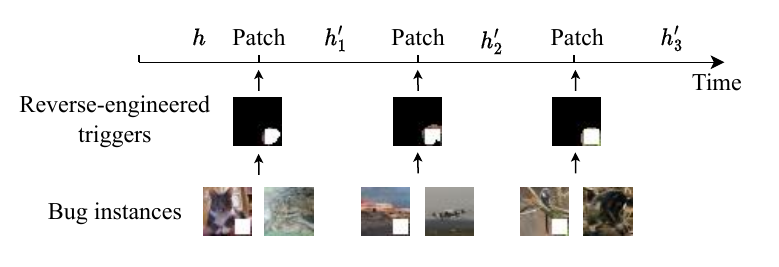}
  \caption{Patching multiple bugs.}
  \label{fig:patch_sequential}
  \vspace{3mm}
\end{figure}

\begin{table}[!t]
  \centering
  \fontsize{7.5}{10}\selectfont
  \vspace{3mm}
  \caption{Patching multiple bugs.}
	\label{tab:impact_patch_sequential}
	{
    \begin{tabular}{|c|c|c|c|c|}
     \hline
    \multirow{2}{*}{\makecell{Foundation model}}   & \multirow{2}{*}{{\makecell{CP}}}   & \multirow{2}{*}{\makecell{Acc}} &  \multirow{2}{*}{\makecell{ASR }}    & \multirow{2}{*}{\makecell{AccB}}     \\
    & & & &\\
    \hline  \hline
    $h$ & $\times$     & 76.46     & 99.82      & 0.10\\  \hline
    $h_1^{\prime}$  &  \checkmark  &      76.51    & 3.43    &   71.43 \\ \hline
    $h_2^{\prime}$  &  \checkmark  &    75.31    & 2.25    &   71.93 \\ \hline
    $h_3^{\prime}$  &  \checkmark  &   74.88    & 2.00    &   71.10\\ \hline
    \end{tabular}%
    }
\end{table}

\myparatight{Patching normal bugs} 
The misclassified input $x_b$ can also be a normal input without a backdoor trigger. \method can be directly applied to patch such bugs. We show the results of patching a normal bug in Table~\ref{tab:general_misclassification}. When $x_b$ is a normal misclassified input, our method still achieves the first two patching goals, while the generalizability goal is not well defined for normal bugs. This is because \method achieves the three patching goals by assuming $x_b$ is a trigger-embedded input from a backdoor attack without distinguishing between
backdoor and normal bugs. 

\begin{table}[!t]
  \centering
  \fontsize{7.5}{10}\selectfont
  \vspace{5mm}
  \caption{Normal vs. backdoor bug.}
	\label{tab:general_misclassification}
	{
    \begin{tabular}{|c|c|c|c|c|c|}
     \hline
    \multirow{2}{*}{\makecell{Bug instance}}    &  \multirow{2}{*}{\makecell{Reverse engineered\\trigger}}  & \multirow{2}{*}{{\makecell{CP}}}   & \multirow{2}{*}{\makecell{Acc}} &  \multirow{2}{*}{\makecell{ASR }}    & \multirow{2}{*}{\makecell{AccB}}     \\
    & & & & &\\
    \hline  \hline
    \multirow{2}{*}{\makecell{Backdoor bug}}  & \multirow{2}{*}{\makecell{\adjustbox{valign=c}{\includegraphics[width=0.03\textwidth]{figs/reverse_triggers/occlusion_trigger.jpg}}}} & \multirow{2}{*}{\checkmark}    & \multirow{2}{*}{76.59}    &  \multirow{2}{*}{2.39}     &    \multirow{2}{*}{73.38}       \\
    &   &   &   &   &   \\\hline
    \multirow{2}{*}{\makecell{Normal bug}}  & \multirow{2}{*}{\makecell{\adjustbox{valign=c}{\includegraphics[width=0.03\textwidth]{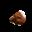}}}} & \multirow{2}{*}{\checkmark}    &  \multirow{2}{*}{76.29}    &  \multirow{2}{*}{-}     &    \multirow{2}{*}{-}       \\
    &   &   &   &   &   \\\hline
    \end{tabular}%
    }
\end{table}

\myparatight{Impact of reference inputs}
A bug instance $(x_b, x_r)$ includes a reference input $x_r$. By default, we assume $x_r$ is a real image from the client. We study whether $x_r$ could be a random input that has a similar feature vector with a real one. Generating such a random input would be easier for the foundation-model provider than the client, who only has black-box access to the foundation model. Therefore, we consider the following scenario: the client sends a misclassified input $x_b$ together with the feature vector of a real reference input, i.e., $h(x_r)$, to the foundation-model provider.  The foundation-model provider  generates a reference input $x_r^{\prime}$ whose feature vector has a high cosine similarity with $h(x_r)$. Specifically, we iteratively update an initial random input $x_r^{\prime}$ for 100 iterations to maximize the cosine similarity between $h(x_r^{\prime})$ and $h(x_r)$ using the Adam optimizer with a learning rate of $1\times10^{-2}$. Figure~\ref{fig:ref_input} in Appendix shows $x_r$ and $x_r^{\prime}$, while Table~\ref{tab:impact_ref_input} shows the results of using $x_r$ or $x_r^{\prime}$ for patching. We have two main observations. First,~\method can successfully reverse engineer the trigger using $x_r^{\prime}$. This is because $x_r^{\prime}$ has a dissimilar feature vector with $x_b$. Second,~\method achieves similar performance when using either a real or random reference input. This is because of the highly similar feature vectors between the real reference input and the random one, and the successfully reverse-engineered trigger. 

\begin{table}[!t]
  \centering
  \fontsize{7.5}{10}\selectfont
  \vspace{3mm}
  \caption{Impact of reference input.}
	\label{tab:impact_ref_input}
	{
    \begin{tabular}{|c|c|c|c|c|c|}
     \hline
    \multirow{2}{*}{\makecell{Reference input}}    &  \multirow{2}{*}{\makecell{Reverse engineered\\trigger}}  & \multirow{2}{*}{{\makecell{CP}}}   & \multirow{2}{*}{\makecell{Acc}} &  \multirow{2}{*}{\makecell{ASR }}    & \multirow{2}{*}{\makecell{AccB}}     \\
    & & & & &\\
    \hline  \hline
    \multirow{2}{*}{\makecell{Real}}  & \multirow{2}{*}{\makecell{\adjustbox{valign=c}{\includegraphics[width=0.03\textwidth]{figs/reverse_triggers/occlusion_trigger.jpg}}}} & \multirow{2}{*}{\checkmark}    & \multirow{2}{*}{76.59}    &  \multirow{2}{*}{2.39}     &    \multirow{2}{*}{73.38}       \\
    &   &   &   &   &   \\\hline
    \multirow{2}{*}{\makecell{Random}}  & \multirow{2}{*}{\makecell{\adjustbox{valign=c}{\includegraphics[width=0.03\textwidth]{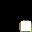}}}} & \multirow{2}{*}{\checkmark}    &  \multirow{2}{*}{76.66}    &  \multirow{2}{*}{2.60}     &    \multirow{2}{*}{72.11}       \\
    &   &   &   &   &   \\\hline
    \end{tabular}%
    }
\end{table}

\myparatight{Impact of other parameters}
We study the performance of our~\method when patching the foundation model and training the downstream classifier using different learning rates, batch sizes, or epochs. We show the results in Table~\ref{tab:parametersettings} in Appendix. We observe that \method achieves the three patching goals across all parameter settings. Our formulated optimization problem in Equation~\ref{eq:loss_all} is a weighted sum of the three loss terms. We further study the impact of the two weights $\lambda_l$ and $\lambda_g$. We show the results in Figure~\ref{fig:impact_lambda}. We have two key observations. First, \method achieves the three goals once the optimization problem incorporates the three loss terms, i.e., $\lambda_l \neq 0$ and $\lambda_g \neq 0$. Second, too large $\lambda_l$ (or $\lambda_g$) may slightly impact the generalizability (or locality) goal. For example, when $\lambda_l \geq 3$ (or $\lambda_g \geq 3$), the AccB (or Acc) slightly decreases. This is because too large $\lambda_l$ (or $\lambda_g$) makes \method minimize more on the locality (or generalizability) loss term than the other two. 

\begin{figure}[!t]
  \centering
  \subfloat[$\lambda_l$]{\includegraphics[width=0.24\textwidth]{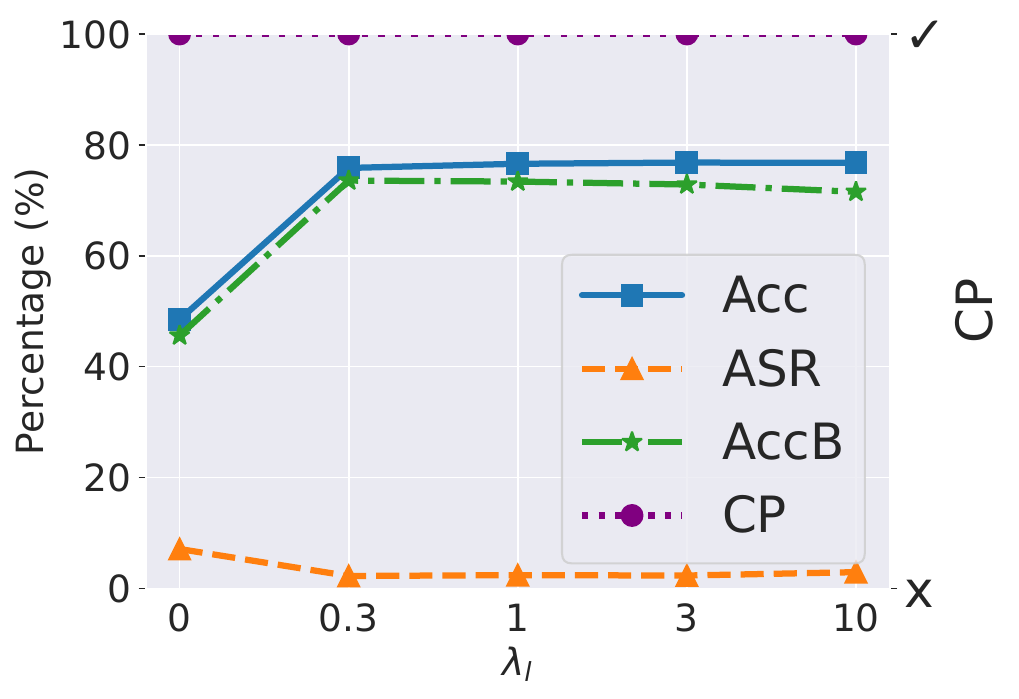}}
  \subfloat[$\lambda_g$]{\includegraphics[width=0.24\textwidth]{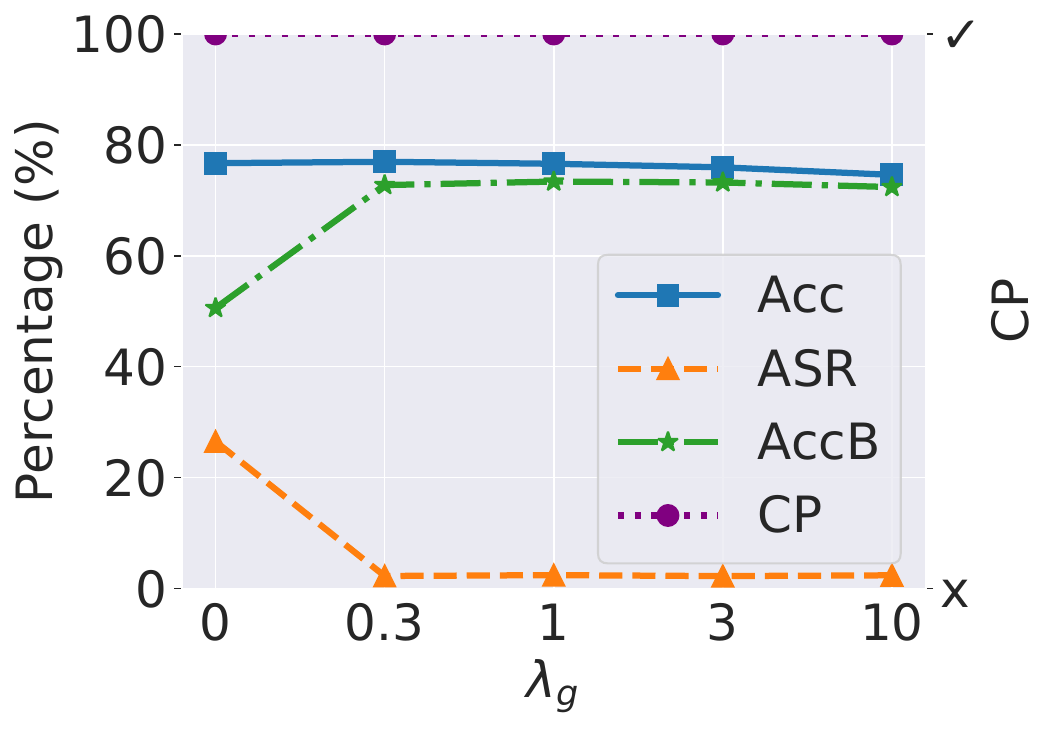}}
  \caption{Impact of $\lambda_l$ and $\lambda_g$.}
  \label{fig:impact_lambda}
\end{figure}

\begin{table}[!t]
  \centering
  \vspace{5mm}
  \fontsize{8}{10}\selectfont
   \caption{Comparing \method with patching downstream classifier alone.}
	\label{tab:baseline_patch_downstream_classifier}
	{
    \begin{tabular}{|c|c|c|c|c|}
     \hline
    \multirow{2}{*}{\makecell{Patching method}}   & \multirow{2}{*}{\makecell{CP}}   & \multirow{2}{*}{\makecell{Acc}}   &  \multirow{2}{*}{\makecell{ASR }}    & \multirow{2}{*}{\makecell{AccB}}     \\
    & & & &\\
    \hline  \hline
    {\makecell{Before patching}}  &  $\times$     & 76.46   & 99.82      & 0.10     \\  \hline \hline
    {\makecell{FT}}  &     \checkmark     &   34.86   &  0.00   &   11.11  \\  \hline
    {\makecell{FT with $\mathcal{D}{train}$}}  &   \checkmark     & 76.54    & 0.44    &   11.14     \\  \hline
    {\makecell{\hongbin{AI-Lancet~\cite{zhao2021ai}}}}  &   \hongbin{\checkmark }    &      \hongbin{74.54}    & \hongbin{0.82}    &   \hongbin{11.31} \\  \hline \hline
    {\makecell{BEAGLE~\cite{cheng2023beagle} + reversed trigger}}  &   \checkmark     &   76.34    & 5.22    &   54.33  \\  \hline
    {\makecell{BEAGLE~\cite{cheng2023beagle} + exact trigger }}  &   \checkmark     &      76.31    & 3.47    &   58.91 \\  \hline 
    {\makecell{\methodtight}}  &  \checkmark   &76.59    & 2.39    &   73.38  \\  \hline
    \end{tabular}%
    }
    \vspace{-2mm}
\end{table}

\subsection{Patching Downstream Classifiers Alone}
\label{sec:patchingdownstream}
Given a misclassified input $x_b$, a client could try patching its downstream classifier directly. We consider the following  methods to patch a downstream classifier  while keeping the foundation model $h$ unchanged.
 
    \myparatight{FT}  FT uses $(x_b, y_b)$ to fine-tune the downstream classifier, where $y_b$ is the true label of $x_b$. In particular, FT obtains a post-patching downstream classifier $f^{\prime}$  by solving the optimization problem:  $\min_{f^{\prime}}  \ell_{CE}(x_b, y_b;f^{\prime}\bigodot h)$, where $\ell_{CE}$ is the cross-entropy loss and $h$ is not changed. 
    
    \myparatight{FT with $\mathcal{D}_{train}$} Since the client has a downstream training dataset $\mathcal{D}_{train}$, the client can fine-tune the downstream classifier using $\mathcal{D}_{train}$ together with the misclassified input $x_b$. 
    In particular, the client uses $\mathcal{D}_{train} \cup \{(x_b, y_b)\}$ to fine-tune the downstream classifier. 

    \hongbin{\myparatight{AI-Lancet~\cite{zhao2021ai}} AI-Lancet is a patching method for classifiers. Given a misclassified input (i.e., $x_b$ in our case) and a correctly classified input (i.e., $x_r$ in our case), AI-Lancet locates the backdoor neurons   and adjusts them to remove the backdoor.  We apply AI-Lancet to patch downstream classifier. We use their neuron-flip variant that achieves better results. }

    \myparatight{BEAGLE~\cite{cheng2023beagle}} BEAGLE reverse engineers a trigger from \emph{multiple} misclassified trigger-embedded inputs and fine-tunes a classifier using  clean training inputs embedded with the reverse engineered trigger and correct labels. However, in our problem setup, we consider a client who detects a single misclassified trigger-embedded input. Therefore, their method of reverse engineering a trigger is not applicable. Instead, we use the trigger reverse engineered by the foundation-model provider using our method  or the exact trigger when applying BEAGLE to patch a downstream classifier.  We use `BEAGLE + reversed trigger' and `BEAGLE + exact trigger' to denote these two cases, respectively. Note that BEAGLE + exact trigger gives an advantage to BEAGLE. 

For all these methods \hongbin{except AI-Lancet, which does not require fine-tuning}, we fine-tune a downstream classifier for 50 epochs with a learning rate of 0.001 using Adam optimizer.  Table~\ref{tab:baseline_patch_downstream_classifier} compares \method with patching downstream classifiers alone. We observe that all patching methods achieve the effectiveness goal. However, FT does not achieve the locality nor generalizability goal; FT with $\mathcal{D}_{train}$ \hongbin{and AI-Lancet} do not achieve the generalizability goal since 
\hongbin{ASRs are low but AccBs are also low,} 
i.e., trigger-embedded inputs are still misclassified as non-target classes; and  BEAGLE  better achieves the generalizability goal than the other baselines, but it still has higher ASR and much lower AccB than \methodtight. We note that BEAGLE + reversed trigger achieves comparable performance with BEAGLE + exact trigger, which further indicates that our method can reverse engineer a  high-quality trigger from a single bug instance.  

\subsection{Adaptive Backdoor Attacks}

A key component of \method is to reverse engineer the trigger from a bug instance. Therefore, we consider adaptive  attacks that aim to enhance the complexity and stealthiness of the trigger, making it more difficult for \method to reverse engineer it. A trigger consists of two dimensions: pattern and location. A pattern can be further characterized by its shape, value, and size. For instance, a pattern could be a square (shape) of 10 $\times$ 10 pixels (size), where each pixel has a value of 1 (value). Location could be a fixed location at the bottom right corner or a random location for each input. 

\begin{table}[!t]
  \centering
  \fontsize{8}{10}\selectfont
  \vspace{5mm}
  \caption{Patching results of \method when a backdoor attack uses different trigger patterns.}
	\label{impact_trigger_pattern}
	{
    \begin{tabular}{|c|c|c|c|c|c|}
     \hline
    \multirow{2}{*}{\makecell{Trigger pattern}}   & \multirow{2}{*}{\makecell{Patching}}    & \multirow{2}{*}{\makecell{CP}}   & \multirow{2}{*}{\makecell{Acc}} &  \multirow{2}{*}{\makecell{ASR }}    & \multirow{2}{*}{\makecell{AccB}}     \\
    &   &   &   &   &   \\
    \hline  \hline
    \multirow{2}{*}{\makecell{\adjustbox{valign=c}{\includegraphics{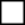}}}}&  Before  & $\times$ &  76.46     & 99.82     & 0.10      \\  \cline{2-6}
    &  After  & \checkmark &    76.59     &2.39    & 73.38        \\  \hline
    \multirow{2}{*}{\makecell{\adjustbox{valign=c}{\includegraphics{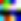}}}}&  Before  & $\times$ &  76.49    & 100.00     & 0.00      \\  \cline{2-6}
    &  After  & \checkmark  & 76.14    & 3.58    &   71.57    \\  \hline
    \multirow{2}{*}{\makecell{\adjustbox{valign=c}{\includegraphics[]{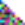}}}}&  Before  & $\times$  &   76.25    &   98.62    &   0.76     \\  \cline{2-6}
    &  After  &  \checkmark &   76.62    & 2.40    &   73.46  \\  \hline
    \multirow{2}{*}{\makecell{\adjustbox{valign=c}{\includegraphics[]{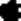}}}}&  Before  & $\times$  &   74.90    & 97.29    &   1.43       \\  \cline{2-6}
    &  After  & \checkmark  & 76.61    & 2.42    &   73.69 \\  \hline
    \multirow{2}{*}{\makecell{\adjustbox{valign=c}{\includegraphics[width=5.5mm]{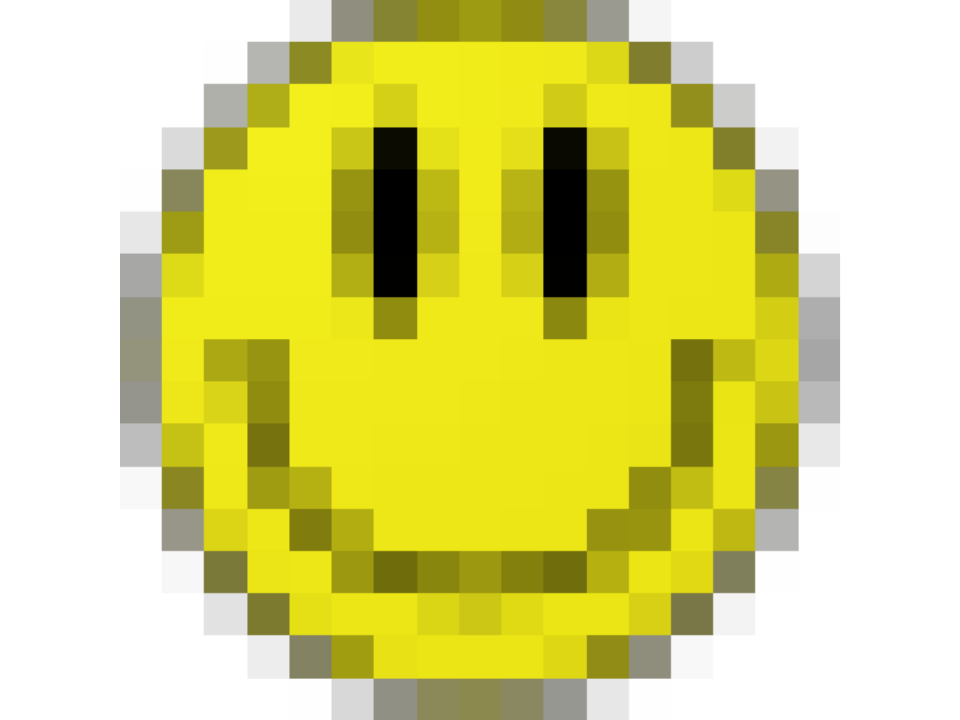}}}}&  Before  & $\times$  &     75.11    & 99.44    &   0.14    \\  \cline{2-6}
    &  After  & \checkmark  &  75.88    &  3.66    &  73.21   \\  \hline
    \multirow{2}{*}{\makecell{\adjustbox{valign=c}{\includegraphics[width=3.5mm]{figs/exp_trigger_pt_rand_1.pdf}}  + \adjustbox{valign=c}{\includegraphics[width=3.5mm]{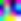}}}} & Before & $\times$ & 76.62    & 100.00    &   0.00 \\  \cline{2-6}
    &  After  & \checkmark &  76.49    & 3.78    &   67.58      \\  \hline
    \end{tabular}
    }
     \vspace{-3mm}
\end{table}

\myparatight{Adapting shapes and values of the trigger pattern} Table~\ref{impact_trigger_pattern} shows the patching results of our method when a backdoor attack uses  trigger patterns with different shapes and values: a $10\times 10$ square with white pixels, a $10\times 10$ square with random pixel values, a $10\times 10$ triangle with random pixel values, an apple logo, an emoji, and a spread-out trigger. The spread-out trigger comprises two $10\times 10$ squares with random pixel values placed diagonally, with the first located at the upper left corner and the second located at the lower right corner. The apple logo and emoji could represent physical objects~\cite{wenger2021backdoor} in physical backdoor attacks.  

Our results indicate that, regardless of the shapes and values, the backdoor attacks are effective at achieving high ASRs before patching. However,  \method can still achieve the three patching goals.    For instance, we observe substantial decreases in ASRs across all trigger patterns; and  AccB remains close to Acc.  Note that AccB is 10\% lower than Acc for the spread-out trigger.  This may be due to the spread-out trigger occupying a larger portion of the image, potentially obscuring key objects and leading to misclassification.

\myparatight{Adapting trigger sizes} 
Figure~\ref{fig:impact_trigger_size} in Appendix shows the patching results of our \method when a backdoor attack uses different trigger sizes, where the trigger pattern is a white square located at the bottom right corner. As the trigger size increases, pre-patching ASR  also increases, which indicates that a backdoor attack with a larger trigger is more effective. However, after patching,  ASR substantially drops and AccB remains close to Acc, which shows that our \method consistently achieves the generalizability goal. Moreover,  our \method also consistently achieves the effectiveness and utility goals across all trigger sizes studied. 

\myparatight{Random trigger location} In our \methodtight, the generalizability loss (Equation~\ref{eq:loss_generalizability}) embeds the reverse-engineered trigger into each input in the validation dataset at the same location. This is because a backdoor attack embeds the trigger at the same, fixed location of inputs.  Therefore, an adaptive backdoor attack is to embed the trigger at a random location of an input. \hongbin{To patch such backdoor bugs, we define a variant of \methodtight, denoted as \methodtight-RL, as follows}: we embed the reverse-engineered trigger at a random location of each input in the validation dataset when defining the  generalizability loss. Table~\ref{tab:impact_trigger_location} shows the patching results for  backdoor attacks with a fixed or random trigger location. Our results show that both \method and \methodtight-RL achieve the three patching goals when the trigger location is fixed. For random trigger location, both achieve the effectiveness and locality goals, but \methodtight-RL better achieves the generalizability goal. 

\begin{table}[!t]
  \centering
  \fontsize{8}{10}\selectfont
  \vspace{3mm}
  \caption{Patching results of our~\method and variant~\methodtight-RL when a backdoor attack uses a fixed or random trigger location.}
    \label{tab:impact_trigger_location}
	{
    \begin{tabular}{|c|c|c|c|c|c|}
     \hline
    \multirow{2}{*}{\makecell{Trigger\\location}}   & \multirow{2}{*}{\makecell{Patching}}    & \multirow{2}{*}{\makecell{CP}}   & \multirow{2}{*}{\makecell{Acc}} &  \multirow{2}{*}{\makecell{ASR }}    & \multirow{2}{*}{\makecell{AccB}}     \\
    & & & & &\\
    \hline  \hline
    \multirow{3}{*}{\makecell{{Fixed}}}&  Before  & $\times$ &  76.46    & 99.82    &   0.10      \\  \cline{2-6}
    &  \method  & \checkmark &    76.59     &2.39    & 73.38        \\  \cline{2-6}
    &  \methodtight-RL  & \checkmark &  75.85    & 3.19    &   71.92   \\  \hline
    \multirow{3}{*}{\makecell{{Random}}}&  Before  & $\times$  &   76.51    & 100.00    &   0.00      \\  \cline{2-6}
    &  \method  &  \checkmark  &  76.60    & 41.43    &   28.42     \\  \cline{2-6}
    &  \methodtight-RL  &  \checkmark &  76.36    & 3.50    &   65.26             \\  \hline
    \end{tabular}%
    }
\end{table}

\myparatight{Source-specific backdoor attacks} Our generalizability loss assumes that any input embedded with the trigger would activate the backdoor. Therefore, one adaptive backdoor attack is the so-called source-specific backdoor~\cite{wang2019neural, gao2019strip, tang2021demon}, in which the backdoor is activated only when the trigger-embedded input is from a particular class called \emph{source class}. Existing source-specific backdoor attacks were designed for classifiers. We extend BadEncoder as a source-specific backdoor attack to foundation models. Due to limited space, we show the technical details about this extension in Appendix~\ref{appendix-source-specific}. 
Table~\ref{tab:one_to_one_attack} shows the patching results, where ASR-source (or ASR-other) is the fraction of trigger-embedded inputs from the source class (or non-source classes) that are classified as the target class by the downstream classifier.  Our results show that ASR-source is much higher than ASR-other before patching (i.e., source-specific backdoor is effective), but \method still successfully achieves the three patching goals.  

\begin{table}[!t]
  \centering
  \fontsize{8}{10}\selectfont
  \vspace{3mm}
  \caption{Patching results for source-specific backdoor attack.}
    \label{tab:one_to_one_attack}
	{
    \begin{tabular}{|c|c|c|c|c|c|}
     \hline
    \multirow{2}{*}{\makecell{Patching}}    & \multirow{2}{*}{\makecell{CP}}   & \multirow{2}{*}{\makecell{Acc}} &  \multirow{2}{*}{\makecell{ASR-source }} & \multirow{2}{*}{\makecell{ASR-other }}    & \multirow{2}{*}{\makecell{AccB}}     \\
    & & & & &\\
    \hline  \hline
    Before  & $\times$ &  75.33    & 64.38  & 37.33    &   10.78     \\  \hline
    After &  \checkmark     &   75.19   &  3.79    &  3.22   &   70.94      \\  \hline
    \end{tabular}
    }
\end{table}

\hongbin{\myparatight{Dynamic backdoor attacks} A dynamic backdoor attack~\cite{salem2022dynamic}  uses multiple triggers and random locations. These attacks were designed for classifiers, and we extend them to foundation models using BadEncoder.  Specifically,  we use two  $10\times10$  triggers, where trigger $t_1$ is blue and trigger $t_2$ is red. When embedding a trigger into an image, the attacker randomly selects one of these triggers and embeds it into the image at a random location.  Suppose a client detects a bug instance containing trigger $t_1$; the foundation-model provider patches the foundation model based on the bug instance; then a client detects another bug instance containing trigger $t_2$; and the foundation-model provider further patches the foundation model. Table~\ref{tab:impact_patch_sequential_dynamic} shows the patching results  using \methodtight-RL. Before patching, the dynamic backdoor attack is very successful.  After patching the first bug instance, \methodtight-RL  achieves the first two goals, but not the generalizability goal. This is because the backdoor attack is still successful when trigger $t_2$ is used. After patching the second bug instance, \methodtight-RL achieves the three goals. In general, when a dynamic backdoor attack uses $n$ different triggers, \methodtight-RL requires $n$ bug instances corresponding to the $n$ triggers to fully remove the backdoor. }

\begin{table}[!t]
  \centering
  \fontsize{7.5}{10}\selectfont
  \vspace{3mm}
  \caption{\hongbin{Patching results of \methodtight-RL against dynamic backdoor attacks.}}
	\label{tab:impact_patch_sequential_dynamic}
	\hongbin{
    \begin{tabular}{|c|c|c|c|c|}
     \hline
    \multirow{2}{*}{\makecell{Foundation model}}   & \multirow{2}{*}{{\makecell{CP}}}   & \multirow{2}{*}{\makecell{Acc}} &  \multirow{2}{*}{\makecell{ASR }}    & \multirow{2}{*}{\makecell{AccB}}     \\
    & & & &\\
    \hline  \hline
    Before patching  &  $\times$   & 76.46     & 97.31      & 0.21\\  \hline
    After patching first bug instance   &  \checkmark  &      76.07    & 38.42    &   35.23 \\ \hline
    After patching second bug instance  &  \checkmark  &   75.16    & 2.32    &   70.48  \\ \hline
    \end{tabular}
    }
\end{table}

\section{Discussion and Limitations}
\label{sec:discussion}

\hongbin{\myparatight{Patching latent-space backdoor attacks} 
\method considers standard backdoor attacks with localized, universal triggers that can be easily implemented in the physical world. Latent-space backdoor attacks~\cite{doan2021backdoor} use a whole-image imperceptible perturbation as a trigger. We extend such backdoor attacks to foundation models as follows: we craft a whole-image perturbation whose  $\ell_\infty$-norm = 0.01 as a  trigger, and then we use BadEncoder to inject this trigger into a  foundation model. 
Table~\ref{tab:latent_space} shows the patching results, where \method + exact trigger means that \method uses the exact trigger instead of the reverse-engineered one for patching. 

We have several observations. First, the attack is highly effective with a high ASR before patching. Second, \method achieves the first two goals, but not the generalizability goal. This is because the interpretation machine learning method cannot reverse engineer the whole-image trigger, as shown in Figure~\ref{fig:latent_triggers}.   Third, \method + exact trigger achieves the three goals, which further confirms that the ineffectiveness of \method at achieving the generalizability goal is because the trigger cannot be accurately reverse-engineered. However, when the whole-image trigger can be reverse-engineered (e.g., by a method designed in the future), our \method can be used to patch the backdoor.  
}

\myparatight{Patching adversarial-example bugs}
\method focuses on patching backdoor bugs and, as a side effect, can also patch normal misclassification bugs. Adversarial examples~\cite{szegedy2013intriguing,carlini2017towards} are another category of bugs for classifiers. Specifically, an attacker can craft a small perturbation $\delta$ such that $x+\delta$ is misclassified, i.e., $f \bigodot h(x)\neq f \bigodot h(x + \delta)$. Suppose the misclassified input $x_b$ in a bug instance is an adversarial example. If only the effectiveness and locality goals are desired, the bug instance can be treated as a normal misclassification bug, and thus our \method can patch the foundation model to fix it. It is an interesting future work to explore how to define the generalizability goal for adversarial-example bugs and adapt \method to achieve it. For instance, one way to define the generalizability goal is that all adversarial examples generated by the same attack  that generated $x_b$ should be correctly classified after patching. Another way is that all adversarial examples whose perturbation sizes are bounded by that in $x_b$ should be correctly classified after patching.

\myparatight{Malicious client} We assume a bug instance is sent from a benign client. We acknowledge that a malicious client may send carefully crafted bug instances to a foundation-model provider with a goal to disrupt the patching process. We believe it is an important future work to study whether and how a malicious client can subvert the patching process via malicious bug instances. 

\begin{table}[!t]
  \centering
  \vspace{5mm}
  \fontsize{8}{10}\selectfont
   \caption{\hongbin{Patching latent-space backdoor attacks.}}
	\label{tab:latent_space}
	\hongbin{
    \begin{tabular}{|c|c|c|c|c|}
     \hline
    \multirow{2}{*}{\makecell{Patching method}}   & \multirow{2}{*}{\makecell{CP}}   & \multirow{2}{*}{\makecell{Acc}}   &  \multirow{2}{*}{\makecell{ASR }}    & \multirow{2}{*}{\makecell{AccB}}     \\
    & & & &\\
    \hline  \hline
    {\makecell{Before patching}}  &  $\times$     & 75.17   & 83.14      & 4.10     \\  \hline \hline
    {\makecell{\methodtight}}  &  \checkmark   &75.40    & 59.82    &   18.43  \\  \hline
    {\makecell{\method + exact trigger}}  &  \checkmark   &76.25    & 7.35    &   71.71  \\  \hline
    \end{tabular}%
    }
    \vspace{-2mm}
\end{table}

\begin{figure}[!t]
  \centering
  \subfloat{\includegraphics[width=0.15\textwidth]{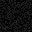}}
  \hspace{4mm}
  \subfloat{\includegraphics[width=0.15\textwidth]{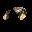}}
  \caption{\hongbin{Visualization of  the exact trigger (\emph{left}) and the trigger (\emph{right}) reverse-engineered by \method in latent-space backdoor attack. 
  }}
  \label{fig:latent_triggers}
\end{figure}

\section{Conclusion and Future Work}
\label{sec:conclusion}

We find that, given a bug instance from a client, a foundation-model provider can patch its foundation model to remove backdoor vulnerabilities without sacrificing its utility. In particular, the provider can patch its foundation model to achieve three goals.  Each goal can be quantified by a loss term, and a foundation model can be patched via minimizing a weighted sum of the three loss terms using a method based on gradient descent. As a side effect, a foundation model can also be effectively patched when a bug instance is for a normal misclassification error. Interesting future works include  patching foundation models to fix adversarial-example bugs and latent-space backdoors, as well as exploring the implications of malicious clients on patching. 

\section*{Acknowledgements}

We thank the anonymous shepherd and reviewers for their constructive comments. This work was supported by NSF under Grant No. 2131859, 2125977, 2112562, 1937786, 1937787, the Army Research
Office under Grant No. W911NF2110182, as well as Ant Group.

\bibliographystyle{plain}
\bibliography{bib}

\appendix

\begin{table*}[!t]
  \centering
  \vspace{3mm}
  \fontsize{8}{10}\selectfont
  \caption{Dataset statistics.}
    \label{tab:data_statistics}
    \begin{tabular}{|c|c|c|c|c|c|}
     \hline
    \multirow{2}{*}{\makecell{Domain}}  &   \multirow{2}{*}{Dataset } & \multirow{2}{*}{\# Training} & \multirow{2}{*}{\# Testing} & \multirow{2}{*}{\# Classes} & \multirow{2}{*}{\makecell{Usage}}     \\
          &       &       &  & & \\
    \hline
    \hline
    \multirow{6}{*}{\makecell{Single-modal Vision}}  & CIFAR10~\cite{krizhevsky2009learning}     &  50,000     &    10,000   & 10   & Pre-training \& Downstream \\
     \cline{2-6}  
    &   STL10~\cite{coates2011analysis}     &   5,000     &  8,000      &  10 & Pre-training \& Downstream     \\
    \cline{2-6}  
    &   ImageNet100-A~\cite{russakovsky2015imagenet}     &   128,420    &   -    & 100   & Pre-training  \\
    \cline{2-6} 
    &   SVHN~\cite{stallkamp2012man}     &    73,257    &    26,032   & 10 & Downstream    \\
    \cline{2-6}  
    &   ImageNet100-B~\cite{russakovsky2015imagenet}     &   126,689    &   5,000    & 100   & Downstream    \\
    \cline{2-6}  
    &   Oxford-IIIT Pets~\cite{parkhi2012cats}     &  3,680    &  3,669     & 37  & Downstream     \\
    \hline
    \multirow{2}{*}{\makecell{Multi-modal Vision}}  &   CLIP-400M~\cite{radford2021learning}    &  $\approx$ 400 Million     &    -   & -    & Pre-training \\
     \cline{2-6}  
    &   CC3M-Sub~\cite{sharma2018conceptual}     &   500,000     &    -    &   -   &  Pre-training   \\
    \hline
    \multirow{3}{*}{\makecell{Language}}    & Wiki103-Sub~\cite{merity2016pointer}     &    250,000    &    -    &  103    &  Pre-training   \\\cline{2-6}  
    &   SST-2~\cite{socher2013recursive}     &   7,792     &    1,821    &   2   &  Downstream   \\\cline{2-6}  
    &   HOSL~\cite{davidson2017automated}     &    8,308    &  2,485      &  2    &  Downstream   \\ \hline
    \end{tabular}
    \vspace{-2mm}
\end{table*}

\begin{table*}[!t]
  \centering
  \fontsize{6}{10}\selectfont
  \vspace{5mm}
  \caption{Pre-training settings of foundation models and training settings of downstream classifiers.}
  \label{tab:combined_table}
  \subfloat[Pre-training settings and backdoor triggers]{
    \label{tab:pre_training_settings}
    \begin{tabular}{|c|c|c|c|c|c|}
     \hline
    \multirow{2}{*}{\makecell{Attack\\method}}     & \multirow{2}{*}{{\makecell{Domain}}}      &  \multirow{2}{*}{\makecell{Backdoor\\trigger}}  & \multirow{2}{*}{\makecell{Pre-training\\algorithm}} &  \multirow{2}{*}{\makecell{Model}} & \multirow{2}{*}{\makecell{Learning\\rate}}   \\
    & & & & &\\
    \hline  \hline
    \multirow{4}{*}{\makecell{BadEncoder}} & \multirow{2}{*}{\makecell{Single-modal\\Vision}} & \multirow{2}{*}{\makecell{\adjustbox{valign=c}{\includegraphics[width=0.03\textwidth]{figs/exp_trigger_pt_white.pdf}}}}     & \multirow{2}{*}{SimCLR~\cite{chen2020simple}}    &  \multirow{2}{*}{ResNet18~\cite{he2016deep}}  &   \multirow{2}{*}{0.001} \\
    &   &   &   &  &  \\\cline{2-6}
    &   \multirow{2}{*}{\makecell{Multi-modal\\Vision}} & \multirow{2}{*}{\makecell{\adjustbox{valign=c}{\includegraphics[width=0.03\textwidth]{figs/exp_trigger_pt_white.pdf}}}}      & \multirow{2}{*}{CLIP~\cite{radford2021learning}}    &  \multirow{2}{*}{ResNet50}   &   \multirow{2}{*}{0.0001}\\
    &   &   &   &  &  \\\hline
    \multirow{4}{*}{\makecell{CorruptEncoder}} & \multirow{2}{*}{\makecell{Single-modal\\Vision}} & \multirow{2}{*}{\makecell{\adjustbox{valign=c}{\includegraphics[width=0.03\textwidth]{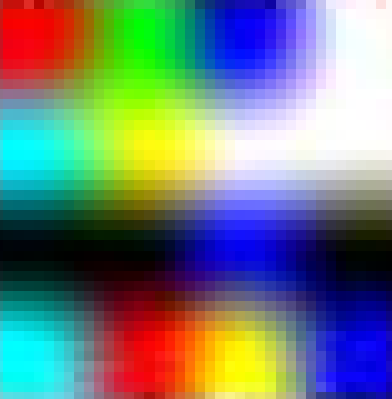}}}}       & \multirow{2}{*}{MoCo v2~\cite{chen2020improved}}    &  \multirow{2}{*}{ResNet18}  &   \multirow{2}{*}{0.001} \\
    &   &   &   &  &  \\\cline{2-6}
    &   \multirow{2}{*}{\makecell{Multi-modal\\Vision}} & \multirow{2}{*}{\makecell{\adjustbox{valign=c}{\includegraphics[width=0.03\textwidth]{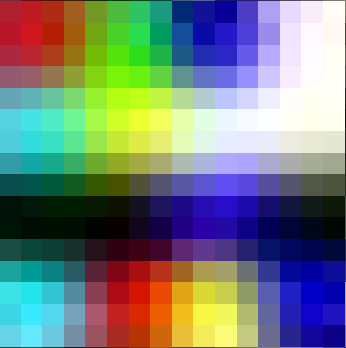}}}}     & \multirow{2}{*}{CLIP}    &  \multirow{2}{*}{ResNet50}   &  \multirow{2}{*}{0.001} \\
    &   &   &   &  &  \\\hline
    \multirow{2}{*}{\makecell{Carlini \& Terzis}}    & \multirow{2}{*}{\makecell{Multi-modal\\Vision}}  & \multirow{2}{*}{\makecell{\adjustbox{valign=c}{\includegraphics[width=0.03\textwidth]{figs/exp_trigger_clip.png}}}}    & \multirow{2}{*}{CLIP}    &  \multirow{2}{*}{ResNet50}  &    \multirow{2}{*}{0.001}\\
    &   &   &   &   & \\\hline
    \multirow{2}{*}{\makecell{POR}}    & \multirow{2}{*}{Language}  & \multirow{2}{*}{\makecell{bb}}    & \multirow{2}{*}{BERT~\cite{devlin-etal-2019-bert}}    &  \multirow{2}{*}{BERT-base} & \multirow{2}{*}{0.0001}  \\
    &   &   &   &   & \\\hline
    \multirow{2}{*}{\makecell{NeuBA}}    & \multirow{2}{*}{Language}  & \multirow{2}{*}{\makecell{$\equiv$}}    & \multirow{2}{*}{BERT}    &  \multirow{2}{*}{BERT-base}   & \multirow{2}{*}{0.0001} \\
    &   &   &   &   & \\\hline
    \end{tabular}
  }
  \subfloat[Training settings of downstream classifiers]{
    \label{tab:downstream_training_settings}
    \begin{tabular}{|c|c|c|c|c|}
     \hline
    \multirow{2}{*}{\makecell{Attack\\method}}     & \multirow{2}{*}{{\makecell{Domain}}}      &  \multirow{2}{*}{\makecell{\# Fully connected\\layers }}  & \multirow{2}{*}{\makecell{\# Epochs}} &  \multirow{2}{*}{\makecell{Learning\\rate}}   \\
    & & & & \\
    \hline  \hline
    \multirow{4}{*}{\makecell{BadEncoder}} & \multirow{2}{*}{\makecell{Single-modal\\Vision}} & \multirow{2}{*}{\makecell{2}}     & \multirow{2}{*}{500}    &  \multirow{2}{*}{0.0001}     \\
    &   &   &   &    \\\cline{2-5}
    &   \multirow{2}{*}{\makecell{Multi-modal\\Vision}} & \multirow{2}{*}{\makecell{2}}    & \multirow{2}{*}{500}    & \multirow{2}{*}{0.0001}     \\
    &   &   &   &    \\\hline
    \multirow{4}{*}{\makecell{CorruptEncoder}} & \multirow{2}{*}{\makecell{Single-modal\\Vision}} & \multirow{2}{*}{\makecell{1}}      & \multirow{2}{*}{50}  &  \multirow{2}{*}{0.01}     \\
    &   &   &   &    \\\cline{2-5}
    &   \multirow{2}{*}{\makecell{Multi-modal\\Vision}} & \multirow{2}{*}{\makecell{1}}    & \multirow{2}{*}{50}    & \multirow{2}{*}{0.01}    \\
    &   &   &   &    \\\hline
    \multirow{2}{*}{\makecell{Carlini \& Terzis}}    & \multirow{2}{*}{\makecell{Multi-modal\\Vision}}  & \multirow{2}{*}{\makecell{1}}   & \multirow{2}{*}{50}    &  \multirow{2}{*}{0.01}    \\
    &   &   &   &    \\\hline
    \multirow{2}{*}{\makecell{POR}}    & \multirow{2}{*}{Language}  & \multirow{2}{*}{\makecell{1}}    & \multirow{2}{*}{10}    &  \multirow{2}{*}{0.001}   \\
    &   &   &   &    \\\hline
    \multirow{2}{*}{\makecell{NeuBA}}    & \multirow{2}{*}{Language}  & \multirow{2}{*}{\makecell{1}}    & \multirow{2}{*}{10}    & \multirow{2}{*}{0.001}    \\
    &   &   &   &    \\\hline
    \end{tabular}
  }
\end{table*}

\begin{table}[!t]
  \centering
  \fontsize{7}{10}\selectfont
  \vspace{5mm}
  \caption{Parameter settings for patching.}
  \label{tab:patching_settings}
\begin{tabular}{|c|c|c|c|c|}
     \hline
    {\makecell{Attack\\method}}     & {{\makecell{Domain}}}      &  {\makecell{Validation\\dataset size}}  & {\makecell{Batch size}}    &   \makecell{Learning\\rate}  \\
    \hline  \hline
    \multirow{2}{*}{\makecell{BadEncoder}} & {\makecell{Single-modal vision}} & {\makecell{50,000}}     & {256}     & 0.001    \\ \cline{2-5}
    &   {\makecell{Multi-modal vision}} & {\makecell{50,000}}    & {32}   &  0.00001 \\\hline
    \multirow{2}{*}{\makecell{CorruptEncoder}} & {\makecell{Single-modal vision}} & {\makecell{50,000}}      & {32}     &   0.001    \\\cline{2-5}
    &   {\makecell{Multi-modal vision}} & {\makecell{50,000}}    & {32}     &   0.00001  \\\hline
    {\makecell{Carlini \& Terzis}}    & {\makecell{Multi-modal vision}}  & {\makecell{50,000}}   & {32}     &   0.00001   \\\hline
    {\makecell{POR}}    & {Language}  & {\makecell{100,000}}    & {32}  &   0.00001  \\\hline
    {\makecell{NeuBA}}    & {Language}  & {\makecell{100,000}}    & {32}   &    0.00001   \\\hline
    \end{tabular}
\end{table}

\begin{table}[!t]
    \renewcommand{\arraystretch}{1.2} 
    \centering
    \vspace{5mm}
    \caption{Results of our~\method when patching the foundation model and training the downstream classifier using different learning rates, batch sizes, or epochs.}
    {
    \fontsize{7.5}{10}\selectfont
    \begin{tabular}{|c|c|c|c|c|c|c|}
        \hline
        \makecell{Phase} & \makecell{Parameter} & \makecell{Value}   & \makecell{CP} & \makecell{Acc}   & \makecell{ASR} & \makecell{AccB} \\ \hline
        \multirow{9}{*}{\makecell{Patching}} & \multirow{3}{*}{\makecell{Learning \\Rate}} & $5 \times 10^{-3}$ & \checkmark & 76.36    & 4.15    &   71.11 \\
        \cline{3-7}  
        & & $1 \times 10^{-3}$ & \checkmark  &  76.59     &2.39    & 73.38  \\
        \cline{3-7}  
        & & $5 \times 10^{-4}$ & \checkmark  &  76.06    & 2.69    &   72.25 \\
        \cline{2-7}  
        & \multirow{3}{*}{Batch Size} & 128 & \checkmark  &  76.73     &3.21    & 73.66  \\
        \cline{3-7}  
        & & 256 & \checkmark  &  76.59     &2.39    & 73.38  \\
        \cline{3-7}  
        & & 512 &  \checkmark  &  76.04     &3.72    & 72.44  \\
        \cline{2-7}  
        & \multirow{3}{*}{Epochs} & 50 & \checkmark  &   76.89    & 3.46    &   71.29\\
        \cline{3-7}  
        & & 100 &  \checkmark  &   76.91    & 2.78    &   71.71    \\
        \cline{3-7}  
        & & 200 & \checkmark  &  76.59     &2.39    & 73.38  \\
        \hline
        \multirow{9}{*}{\makecell{Training \\Downstream \\Classifier}} & \multirow{3}{*}{\makecell{Learning \\Rate}} & $1 \times 10^{-3}$ & \checkmark  & 76.71    & 2.56    &   72.31 \\
        \cline{3-7}  
        & & $1 \times 10^{-4}$ &  \checkmark  &  76.59     &2.39    & 73.38  \\
        \cline{3-7}  
        & & $5 \times 10^{-5}$ & \checkmark  & 77.04    & 2.71    &   72.88  \\
        \cline{2-7}  
        & \multirow{3}{*}{Batch Size} & 32  & \checkmark & 77.14    & 2.43    &   73.14 \\
        \cline{3-7}  
        & & 64 & \checkmark  &  76.59     &2.39    & 73.38  \\
        \cline{3-7}  
        & & 128 & \checkmark  & 76.56    & 2.40    &   72.56\\
        \cline{2-7}  
        & \multirow{3}{*}{Epochs} & 100 & \checkmark  &     77.44    & 2.49    &   73.40   \\
        \cline{3-7}  
        & & 300 &  \checkmark  &   76.94    & 2.49    &   72.72  \\
        \cline{3-7}  
        & & 500 & \checkmark  &  76.59     &2.39    & 73.38  \\
        \hline
    \end{tabular}
    }
    \label{tab:parametersettings}
\end{table}

\begin{figure*}[!t]
  \centering
  \subfloat[Impact of trigger size on Acc]{\includegraphics[width=0.32\textwidth]{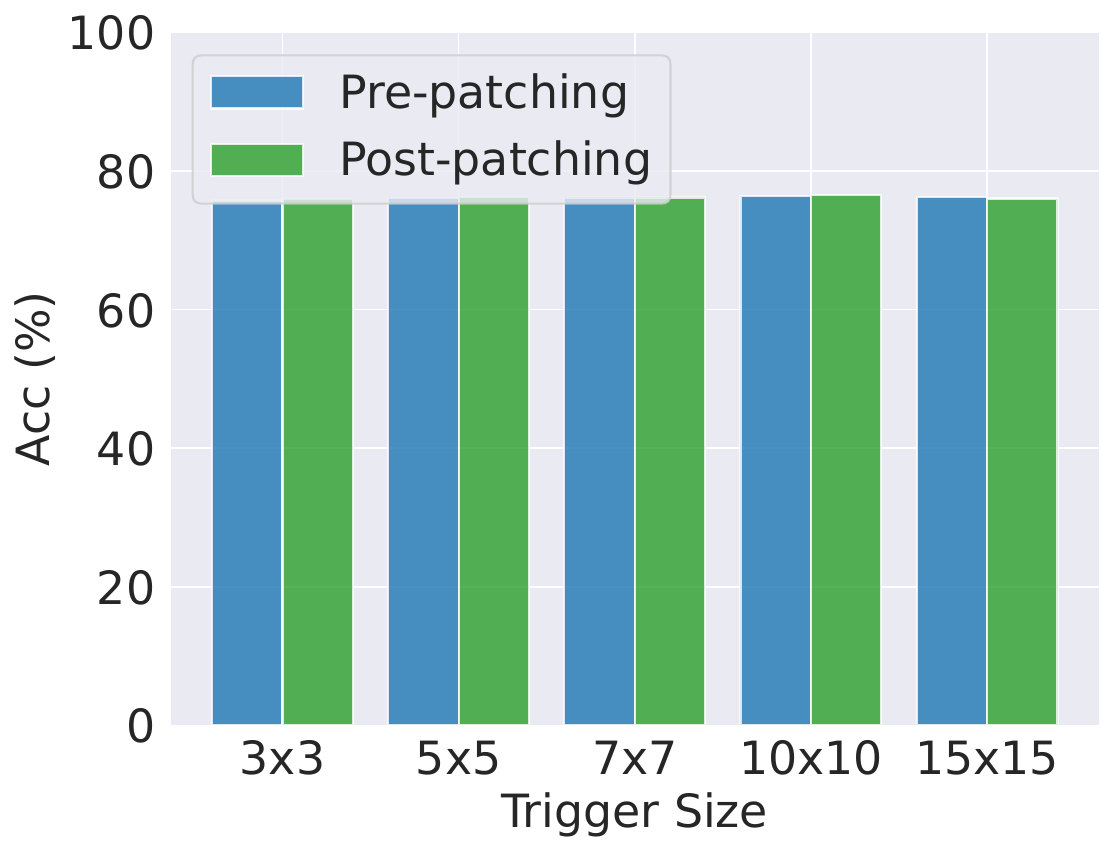}\label{fig:impact_trigger_size_acc}}
  \hfill
  \subfloat[Impact of trigger size on ASR]{\includegraphics[width=0.32\textwidth]{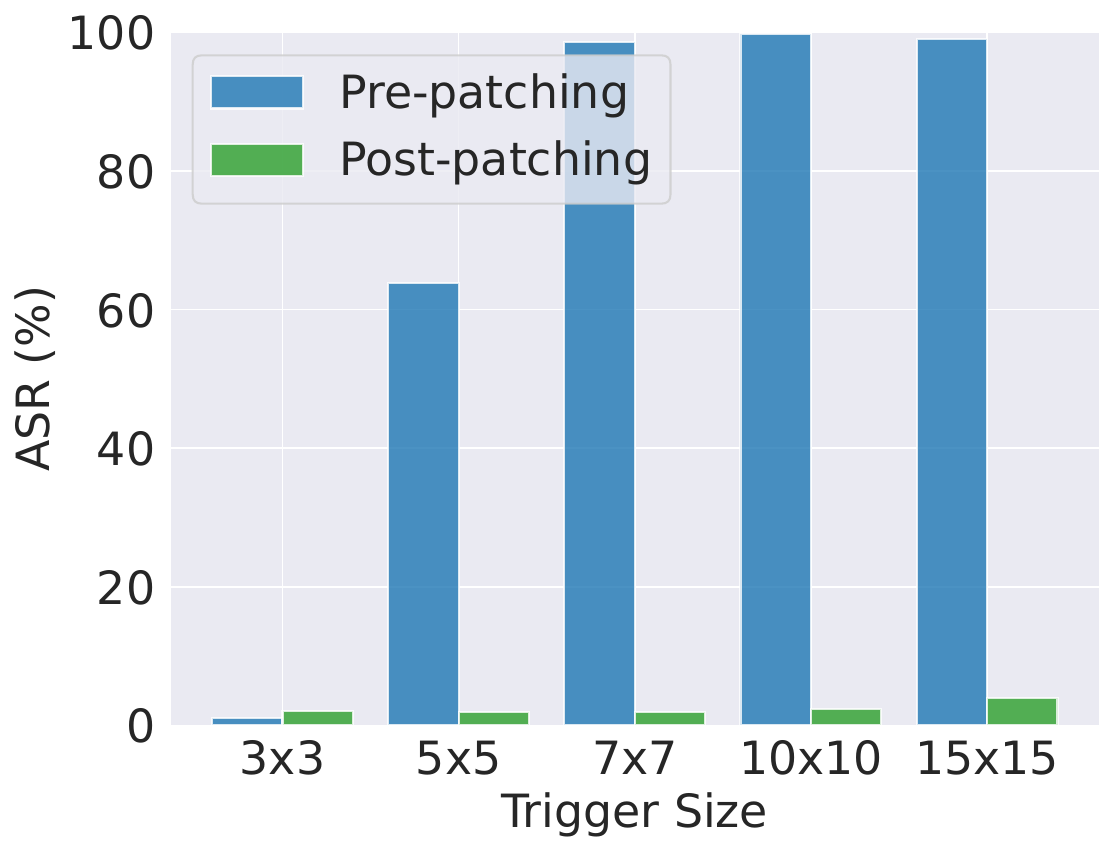}\label{fig:impact_trigger_size_asr}}
  \hfill
  \subfloat[Impact of trigger size on AccB]{\includegraphics[width=0.32\textwidth]{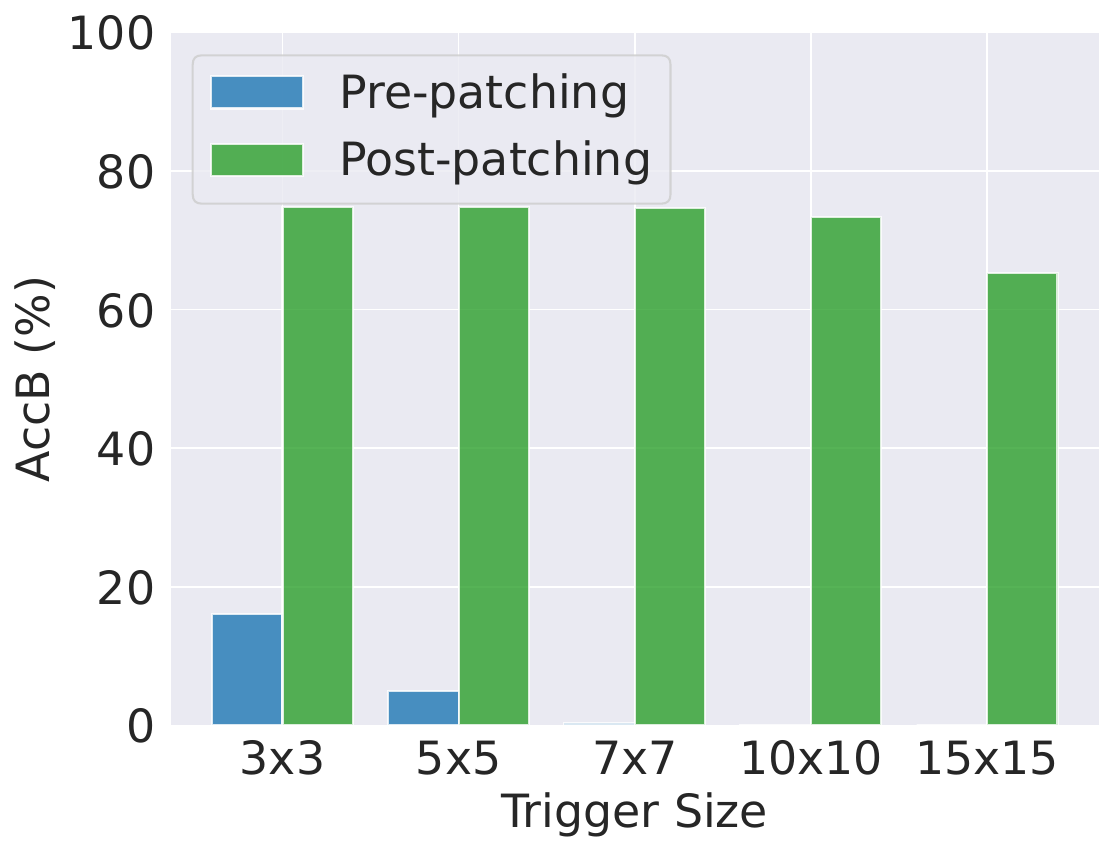}\label{fig:impact_trigger_size_accb}}
  \caption{Patching results of our~\method when a backdoor attack uses different trigger sizes.}
  \label{fig:impact_trigger_size}
\end{figure*}

\section{Source-specific Backdoor Attacks}
\label{appendix-source-specific}
\myparatight{Extending BadEncoder as a source-specific backdoor attack}
\begin{figure}[!t]
  \centering
  \includegraphics[width=0.45\textwidth]{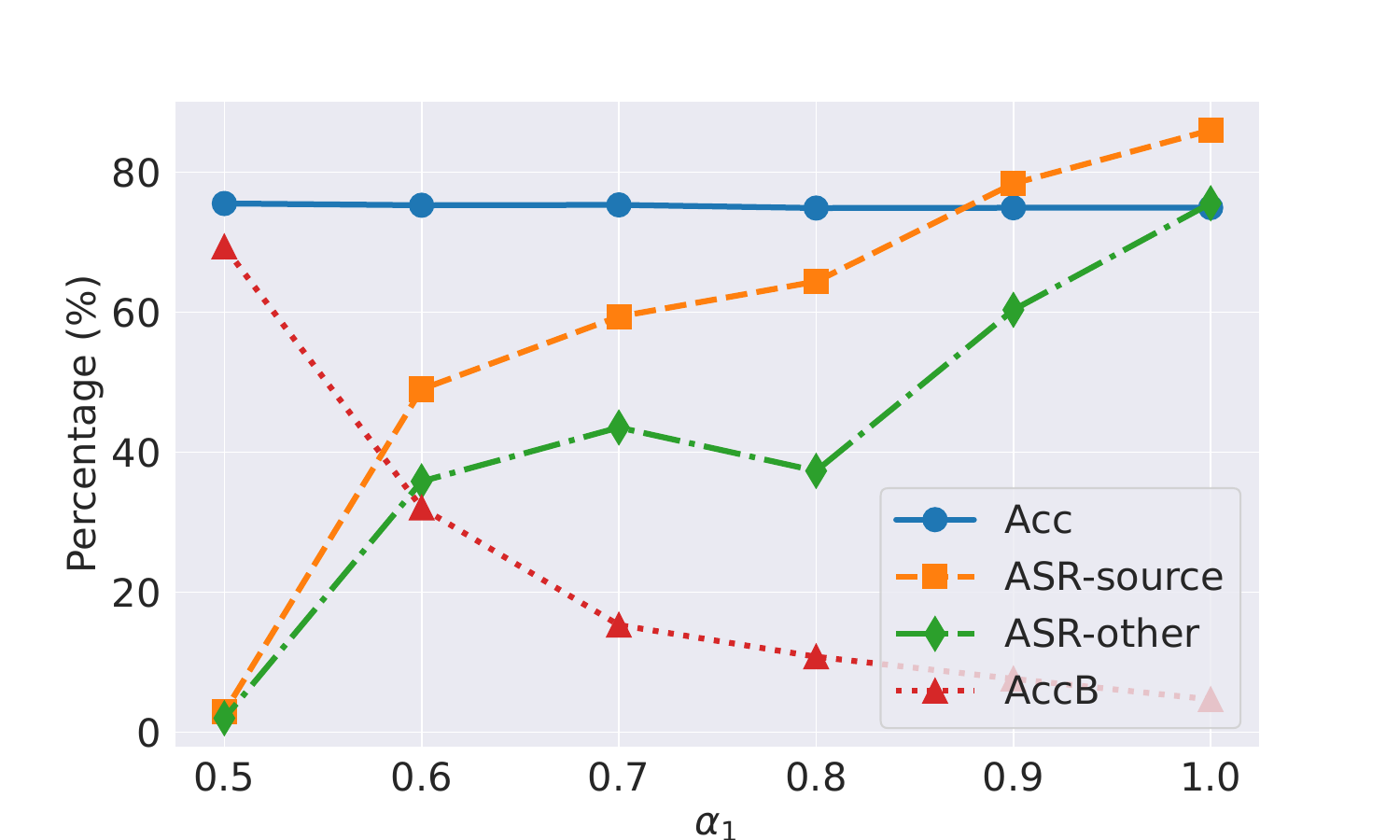}
  \caption{Impact of $\alpha_1$ on the source-specific backdoor attack.}
  \label{fig:source_specific_alpha}
\end{figure}
BadEncoder~\cite{jia2022badencoder} formulates the attacker's effectiveness goal as an effectiveness loss. The effectiveness loss quantifies the similarity between the feature vectors of inputs embedded with the trigger and the feature vector of a reference input. It can be defined as follows:
\begin{align}
\mathcal{L}_0 = - \frac{1}{|\mathcal{D}s|} \sum_{x \in \mathcal{D}s} sim(h(x\oplus t),h(x_r)),
\end{align}
where $\mathcal{D}s$ denotes a shadow dataset used by the attacker, $h$ denotes the backdoored foundation model,  $t$ denotes the trigger, and $x_r$ denotes the reference input. When the effectiveness loss $\mathcal{L}_0$ is optimized, the backdoored foundation model $h$ is likely to output similar feature vectors $h(x\oplus t)$ and $h(x_r)$ for any input $x$.

To achieve the source-specific goal, we introduce the following two main modifications to BadEncoder. First, we give advantage to the attacker by assuming that the attacker has access to the downstream training dataset, which includes labeled data. And the attacker uses this dataset as shadow dataset $\mathcal{D}s$. Second, we split the effectiveness loss $\mathcal{L}_0$ into two losses that are formulated as follows:
\begin{align}
&   \mathcal{L}_{source} =  -  \frac{1}{|\mathcal{D}_{source}|} \sum_{x \in \mathcal{D}_{source}} sim(h(x\oplus t),h(x_r)), \\
&   \mathcal{L}_{other} =  -  \frac{1}{|\mathcal{D}_{other}|} \sum_{x \in \mathcal{D}_{other}} sim(h(x\oplus t),h(x)),
\end{align}
where $\mathcal{D}_{source}$ represents the data from the source class in $\mathcal{D}s$, while $\mathcal{D}_{other}$ represents the data from the other classes. The updated effectiveness loss $\mathcal{L}_0^{\prime} =\alpha_1 \mathcal{L}_{source} + \alpha_2 \mathcal{L}_{other}$, where $\alpha_1$ and $\alpha_2$ balance the weights between the two losses. Intuitively, $\mathcal{L}_{source}$ measures the similarity between the feature vector of any input from the source class embedded with the trigger, and the feature vector of a reference input. However, $\mathcal{L}_{other}$ quantifies the similarity between the feature vector of an input from the other classes  embedded with the trigger, and the feature vector of its clean version.  Therefore, by optimizing these losses, the backdoored foundation model $h$ is more likely to output attacker-desired (or benign) feature vectors when inputs from the source class (or other classes) embedded with the trigger. In other words, the backdoor is more likely to be activated when inputs from the source class are embedded with the backdoor trigger.

\myparatight{Parameter settings}
We use the default parameter settings of BadEncoder but search for the optimal combination of weights $\alpha_1$ and $\alpha_2$ for the attacker. In particular, we explore various values of $\alpha_1$ and $\alpha_2$ within the range of [0, 1], while ensuring $\alpha_1+\alpha_2=1$. Figure~\ref{fig:source_specific_alpha} shows the results of varying $\alpha_1$ values. Note that ASR-source (or ASR-other) is the fraction of trigger-embedded inputs from the source class (or non-source classes) that are classified as the target class by the downstream classifier. Therefore, the attacker's goal is to maximize the gap between ASR-source and  ASR-other. We give advantages to the attacker by using the optimal $\alpha_1=0.8$ and $\alpha_2=0.2$, under the assumption that the attacker has access to the downstream testing dataset.

\begin{figure}[!t]
  \centering
  \subfloat[Real reference input $x_r$]{\includegraphics[width=0.2\textwidth]{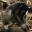}}
  \hfill
  \subfloat[Random reference input $x_r^{\prime}$]{\includegraphics[width=0.2\textwidth]{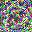}}
  \caption{Visualization of a real reference input and a random reference input.}
  \label{fig:ref_input}
\end{figure}

\end{document}